\documentclass{amsart}

\usepackage{amsfonts}

\usepackage{graphicx}
\usepackage{xcolor}

\def\cal{\mathcal}

\input{tcilatex}
\input tcilatex

\begin{document}

\title[heterogeneous boiling]{How heterogeneous wettability enhances boiling}
\author[J.C. Fernandez Toledano]{Juan Carlos Fernandez Toledano $^{1}$} \author[J. De Coninck] {Jo\"el De Coninck $^{1}$}  
\address{$^{1}$ Laboratoire de Physique des Surfaces et Interfaces\\
Universit\'{e} de Mons, 20 Place du Parc, 7000 Mons, Belgium\\
Emails: joel.deconinck@gmail.com; carlos.toledano@gmail.com}
\author[F. Dunlop]{Fran\c cois Dunlop $^{2}$} \author[T. Huillet]{Thierry Huillet $^{2}$ }  
\address{$^{2}$ Laboratoire de Physique Th\'{e}orique et Mod\'{e}lisation,
CNRS UMR-8089\\ CY Cergy Paris Universit\'{e}, 
95302 Cergy-Pontoise, France\\
E-mails: dunlop@cyu.fr;  huillet@cyu.fr}

\begin{abstract}
For super-heated water on a substrate with hydrophobic patches immersed in a hydrophilic matrix, one can choose the temperature so that micro-bubbles will form, grow and merge on the hydrophobic patches and not on the hydrophilic matrix.
Until covering a patch, making a pinned macro-bubble, a bubble has a contact angle $\pi-\theta_2$, where $\theta_2$ is the receding contact angle of water on the patch material.
This pinned macro-bubble serves as the initial condition of a quasi-static growth process, à la Landau, leading to detachment through the formation of a neck, so long as depinning and dewetting of the hydrophilic matrix was avoided during the growth of the pinned bubble: the bubble contact angle should not exceed
$\pi-\theta_1$, where $\theta_1$ is the receding contact angle of water on the matrix material.
The boiling process may then enter a cycle of macro-bubbles forming and detaching on the patches; the radii of these patches can be optimized for maximizing the heat transfer for a given substrate area.

For this analysis to become quantitative, we revisit the Young-Laplace quasi-static evolution of key physical quantities, such as bubble energy, as functions of bubble growing volume, when gravity is either significative or negligible: this concerns both pinned bubbles on a fixed circular footprint (Dirichlet boundary conditions) and free un-pinned bubbles with a fixed contact angle (Neumann boundary conditions).
\end{abstract}
\maketitle


\section{Introduction}
\begin{figure}[htp!] 
\centering
 \includegraphics[width=\textwidth]{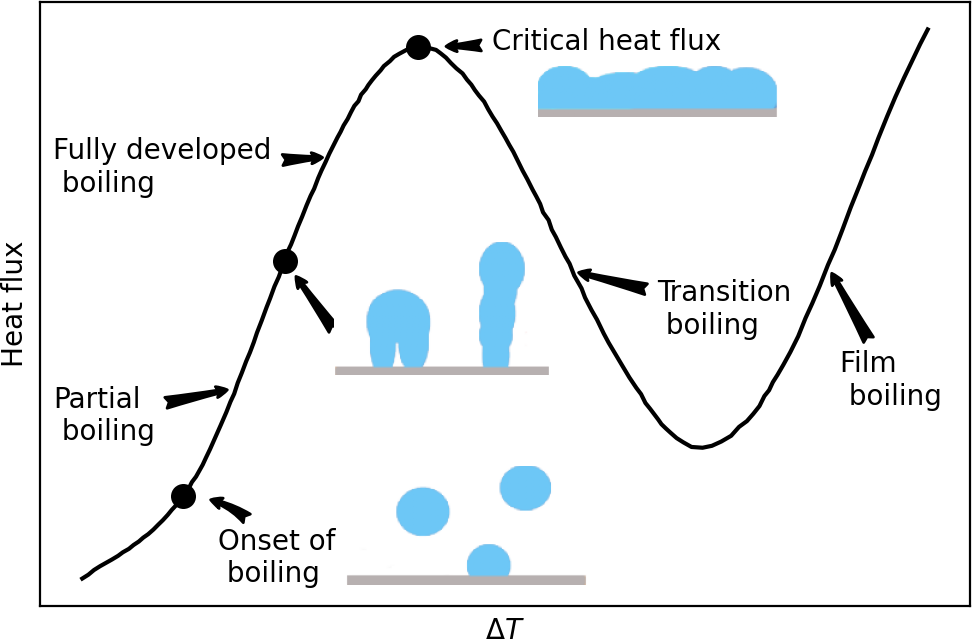} 
\caption{Heat flux versus wall superheat}\label{fig:0}
\end{figure}

Boiling is a common heat transfer technique. It occurs when a liquid is exposed to a surface which is maintained above saturation temperature. This surface can be homogeneous or heterogeneous. The wall temperature ($T_{wall}$) of the liquid needs to be higher than the saturation temperature of the liquid ($T_{sat}$). If the wall temperature is sufficiently high, a thin layer of vapour is formed. This condition, characterized by a vapour film formation on the heated surface, is called film boiling. The extensive study on the effect of the very large superheat difference ($\Delta T:=\Delta T_{sh} = T_{wall}- T_{sat}$) between temperature of the heating surface ($T_{wall}$) and the saturated liquid ($T_{sat}$), was first done by Nukiyama, \cite{Nuki66}. However, it was the experiment by Farber and Scorah, \cite{Farber48}, that gave the complete picture of the heat transfer rate in the pool boiling process (boiling without flow) as a function of $\Delta T_{sh}$. See also \cite{VMC19} and references therein. The various regimes of boiling in a typical case of pool boiling in water at atmospheric pressure are shown in Fig.~\ref{fig:0}. It is the conventional representation of heat flux versus wall superheat. Different ranges can be identified while $\Delta T_{sh}$ is increased. 

This work focuses on pool boiling phenomena, looking in detail at the evolution of a single bubble. When the wall temperature reaches the condition $T_{wall}$=$T_{onb}$, a vapour bubble nucleates on the heater surface. The effect of the wettability on $T_{onb}$ has been studied in Bourdon et al, \cite{Bourdon15} . This paper shows that a thin hydrophobic coating on a surface will significantly lower the onset of boiling. However, at the same time, it will promote the formation of a vapour layer which will in turn affect the critical heat flux.

To avoid this problem, heterogeneous surfaces have been proposed, \cite{VMC19}. These surfaces are generally hydrophilic with hydrophobic patches to decrease the onset of boiling and at the same time increase the critical heat flux. 

The natural question is about the size of these patches. Is there an optimal size to promote boiling? To decrease $T_{onb}$? This question refers of course to the basic mechanism controlling boiling on top of the surface. 

Most probably, nanodroplets of vapour will form everywhere and will coalesce preferentially on top of the hydrophobic patches. This coalescence will occur until the contact area of the patch is covered by vapour and then the vapour bubble will grow until a certain volume where it will detach. 

Herewith we consider the case where boiling includes a stage of slow bubble growth, well approximated by equilibrium shapes. Other cases with ultrafast heat transfer regimes exist but are not considered here; see \cite{SSS15}.

\begin{figure}[htp!] 
 \centering
  \includegraphics[width=\textwidth]{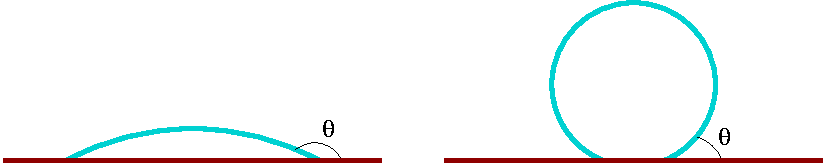} 
 \caption{Vapour bubble on hydrophobic substrate (left) and on hydrophilic substrate (right). 
The angle $\theta$ is the water receding contact angle.}\label{fig:a}
\end{figure}

It has been observed that both hydrophobic substrates and hydrophilic substrates
have advantages with respect to boiling, see \cite{PC09} and \cite{XGL21}.
When the water contact angle $\theta$ tends to 180$^\circ$ (Fig.~\ref{fig:a}, left), the
configuration
approaches a configuration without bubble and the nucleation barrier disappears
as is clear from Young's equation. This is favorable in the first step of the
boiling process. However, as we shall see, the almost flat bubble will need to
grow to a very large volume before developing a neck and detaching. This is
unfavorable.
On the other hand, when $\theta$ tends to zero (Fig.~\ref{fig:a}, right), the configuration
tends to that of a free bubble, homogeneous nucleation with a high barrier.
The presence of a neck from the beginning, a favorable factor, is not enough
to win the game.
A natural answer is to try a patterned surface allowing nucleation on
hydrophobic patches, surrounded by a hydrophilic matrix confining the base of
the bubbles and forcing formation of neck. Whether it works, and with what size
of patch and detaching bubbles, is the main motivation in the present study.

The organization of the paper is as follows:

In Section 2.1, using Young-Laplace equations for axisymmetric bubbles, we
compute the quasi-equilibrium evolution à la Landau of various physical quantities of
interest as a function of its growing volume, first for pinned bubbles on a
fixed circular footprint (Dirichlet boundary conditions). It includes the
mechanical (potential) energy $E$, the contact angle $\alpha $, the height $h
$, the pressure $p$ and the curvature radius $R$ at the apex. The
quasi-equilibrium regime stops when the energy attains its maximal value $%
E_{\max }$ where the bubble enters a non-equilibrium regime leading to its
detachment after the formation of a neck.

For completeness, in Section 2.2, we do the same for free (un-pinned)
bubbles having a fixed contact angle (Neumann boundary conditions). Of
particular interest in this setup is the quasi-equilibrium evolution of the
radius of the bubble footprint till its detachment. We show that this radius
can grow and then shrink in the process.

In the latter two studies, gravity is present and competes with surface
effects.

In Section 2.3, we show how these computations simplify in the absence of
gravity (or when gravity is negligible), leading to spherical cap profiles.
The Young-Laplace equations are explicitly integrable with the evolution of $%
\left( E,h,p\right) ,$ as functions of volume, following directly from the
one of $\alpha .$

We then address the question of the onset of boiling of water on a hot
substrate, through vapour micro-bubble nucleation, so when gravity can be
neglected and when the driving force of bubble growth is a temperature
gradient. We show that, in order for micro-bubbles to form and grow, the
applied $\Delta T=T-373$ must exceed some $\Delta T_{\text{onb}}$ that can
be estimated as a function of the substrate wettability. This $\Delta T_{%
\text{onb}}$ is shown to be a decreasing function of the liquid contact
angle at the base of the bubble and so hydrophobicity favors boiling. If the
substrate is thus made of hydrophobic patches immersed in a hydrophilic sea
(patterning), for each choice of their wettabilities, there is a choice of $%
\Delta T_{\text{onb}}$ for which micro-bubbles form and grow on the
hydrophobic patches and not on the hydrophilic sea. Once these micro-bubbles
have formed on the hydrophobic part of the substrate, each grows, showing up
a hydrophobic liquid contact angle, possibly merge till covering the patch
where the macro-bubble thus formed becomes pinned. At this point, the
contact angle of the liquid at the base of the bubble is the hydrophobic
receding liquid contact angle. This macro-bubble serves as the initial
condition of its quasi-equilibrium evolution described in Section 2.1 till
detachment.

In Section 3, we address the question of the temporal evolution of the
quasi-static bubble growth regime. Using an Onsager ansatz for the evolution
of the number of vapour moles and assuming vapour can be treated as a
perfect gas, we show that during its growth stage, the bubble volume grows
`nearly' linearly with time both in the presence of gravity and when gravity
is negligible.

Section 4 is devoted to a discussion of the `optimal' patch radius $r$ as a
function of the contact angle of the liquid on the patch (hydrophobic) and
the liquid contact angle on the matrix (hydrophilic). By optimal, we mean the radius for which heat transfer is maximal, for a given substrate area and for given materials specifying the contact angles on the patch and the matrix.

\section{Computation of the bubble profiles}

In \cite{LH91} the authors analyze the equilibrium bubble profiles when air is blown quasi-statically through a nozzle. When dealing with boiling of water on a patterned substrate, the driving process for bubble formation is now heat transfer at the circular patch boundary, generating quasi-statically water vapor.  The successive bubble profiles, under the quasi-static approximation, clearly are the same. Pool boiling based on principles of equilibrium thermodynamics was shown relevant in \cite{AGW17}.

\begin{figure}[htp!] 
 \centering
  \includegraphics[width=\textwidth]{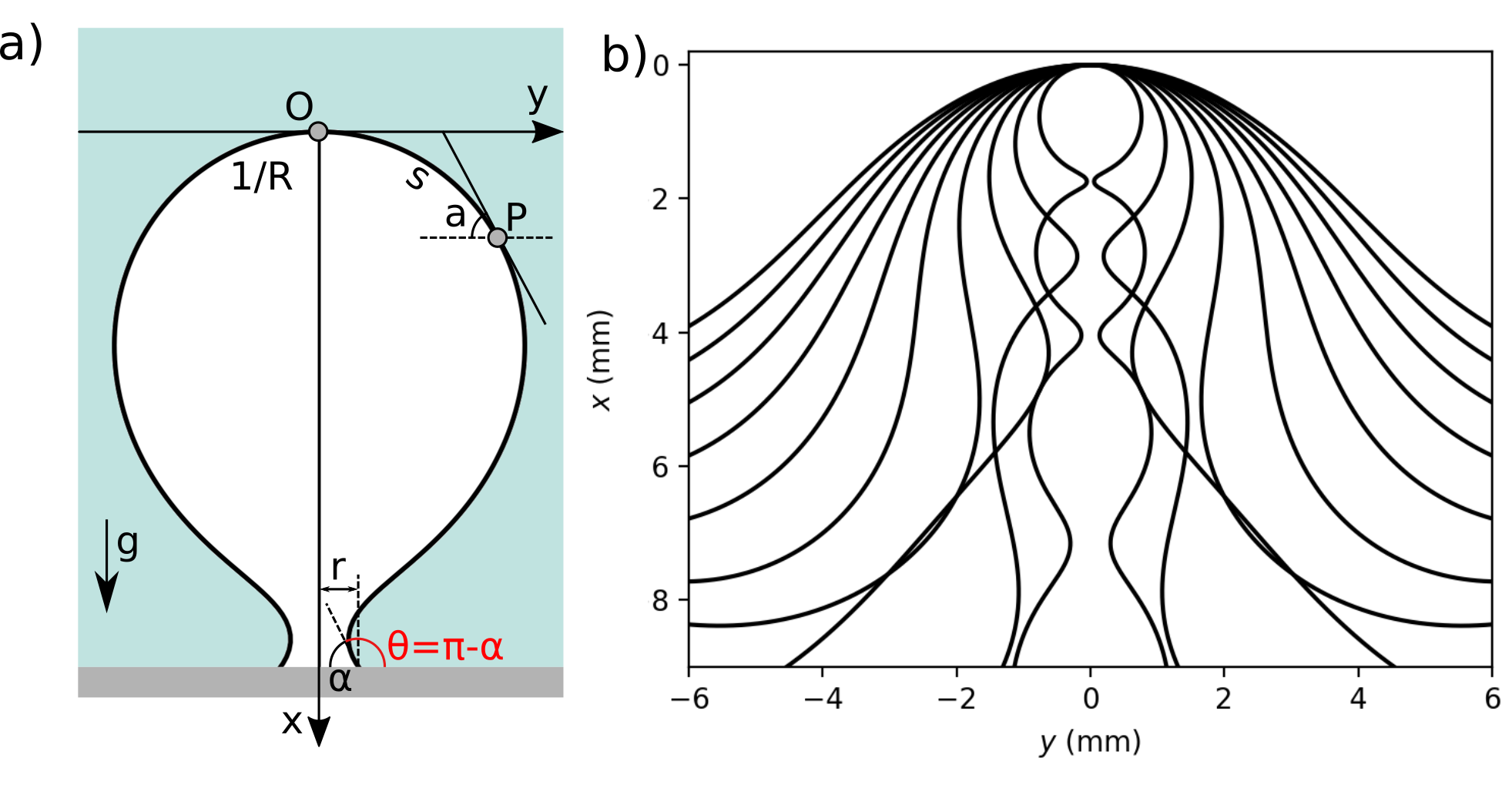} 
 \caption{a) Coordinate system for a gas bubble surrounded by liquid on top of a substrate. b) Family of solutions for Eqs. ~\ref{eq:1} for water at $373^\circ$ K.}\label{fig:1}
\end{figure}

The coordinate system is outlined in Fig.~\ref{fig:1}a with the origin $O$
located at the apex, $Ox$ is the axis of symmetry and with the $y$-axis
radially outwards.

At a point at abscissa $x$ from $O$, the pressure $p_{F}$ in the liquid is
given by 
\begin{equation}
p_{F}=p_{O}+\rho gx
\end{equation}
where $p_{O}$ is the pressure at $O$ in the liquid, $\rho $ is the liquid
density and $g$ the gravity constant. The vapour pressure $p$ in the bubble
is given by the Laplace equation:
\begin{equation}
\Delta p=p_{F}-p=-2\gamma H ,
\end{equation}
where $H=(1/R_1+1/R_2)/2$ is the mean curvature , and $R_{1}$ and $R_{2}$ are the principal radii of curvature with signs, at a point $%
P=\left( x,y\right) $ of the liquid/vapour interface and $\gamma $ the
liquid/vapour surface tension. The vapour pressure $p$ is assumed uniform
inside the bubble due to the comparatively small value of the vapour
density. Therefore, with $\rho $ the water density, 
\begin{equation}
p=p_{O}+\rho gx+\gamma \left( \frac{1}{R_{1}}+\frac{1}{R_{2}}\right) .
\end{equation}
At $x=0$, $R_{1}=R_{2}=:R$ where $R$ is the radius of curvature at the apex $%
O$. Therefore $p=p_{O}+2 \gamma/ R$ and so 
\begin{equation}
\gamma \left( \frac{1}{R_{1}}+\frac{1}{R_{2}}\right) =-\rho gx+\frac{2\gamma 
}{R}
\end{equation}

For axi-symmetric shapes, the radii of curvature $R_{1}$ and $R_{2}$ can be
expressed as 
\begin{eqnarray}
\frac{1}{R_{1}} &=&\frac{da}{ds} \nonumber\\
\frac{1}{R_{2}} &=&\frac{\sin a}{y}
\end{eqnarray}
where $s$ is the profile arc-length from the origin located at the apex to a
given point $P$ of the profile and $a$ the angle between the tangent of the
profile at $P$ and the horizontal (recall $\pi -a$ is the contact angle seen
from the liquid). Then 
\begin{eqnarray}
\gamma \left( \frac{da}{ds}+\frac{\sin a}{y}\right) &=&-\rho gx+\frac{%
2\gamma }{R} \nonumber\\
\frac{da}{ds} &=&-\frac{\rho gx}{\gamma }+\frac{2}{R}-\frac{\sin a}{y}
\end{eqnarray}

We obtain the final system $\left( S\right) $ of ordinary differential
 equations (LH for short) with three $s-$dependent variables $\left( x,y,a\right) $, see \cite{LH91}, %
\begin{eqnarray}\label{eq:1}
\frac{dx}{ds} &=&\sin a \nonumber\\
\frac{dy}{ds} &=&\cos a \nonumber\\
\frac{da}{ds} &=&-\frac{\rho gx}{\gamma }+\frac{2}{R}-\frac{\sin a}{y}
\end{eqnarray}
with initial conditions $x(0)=y(0)=a(0)=0.$ This system $\left( S\right) $
can be integrated numerically to generate the bubble profile for various
values of the control parameter $R$. In the boiling experiment of water, we
take $\rho =959$ kg/m$^{\text{3}}$ and $\gamma =0.0588$ N/m at $373$ ${%
{}^{\circ }}$K$.$

We can create the profiles computed as functions of $s$ for different
values of the inverse curvature $R$ at the apex. The profiles are Delaunay
curves. 
The profile curves $y=y\left( x\right) $ are then available as well, \cite{LH91}. The family of profile solutions to the (say LH for Longuet-Higgins) Eqs.~\ref{eq:1} is shown in Fig.~\ref{fig:1}b.

The latter system $\left( S\right) $ could be adimensionalized by
introducing the capillary length as the length-scale $\left( \gamma /\left( \rho g\right) \right)
^{1/2}=2.453$ mm. 


\subsection{Pinned bubble on a patch with prescribed radius at the bottom $%
y_{B}:=r$ (Dirichlet boundary condition)}

\begin{figure}[htp!] 
 \centering
 \includegraphics[width=\textwidth]{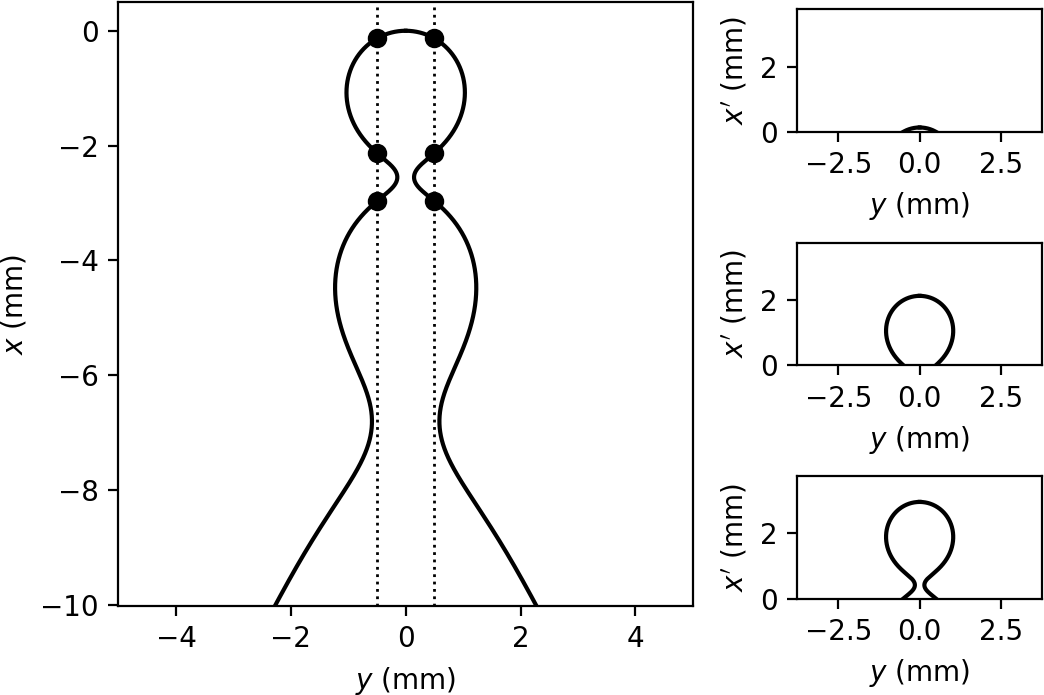} 
 \caption{Profile computed for water at $T=373$ K for a curvature radius at the apex of $R=0.98$ mm. The dotted vertical lines represent a patch radius at $y=y_B=0.49$ mm that intersect the profile at three $x$ values. The corresponding profiles are shown in the three right subplots.}\label{fig:2}
\end{figure}

To obtain the solutions for a given contact patch with radius $y_{B}$, we
intersect the profiles computed in the previous figure with the lines $y=\pm
y_{B}$. For example, for a radius of curvature $R=0.98$ mm, there are three curves that
satisfy $y_B=0.49$ mm as is shown in Fig.~\ref{fig:2}.

The system $\left( S\right) $ yields the corresponding full bubble profile
pinned on the patch, in particular the value $h:=x_{B}$ of the height of the
bubble with patch radius $r:=y_{B}$. 

\begin{figure}[htp!] 
 \centering
 \includegraphics[width=\textwidth]{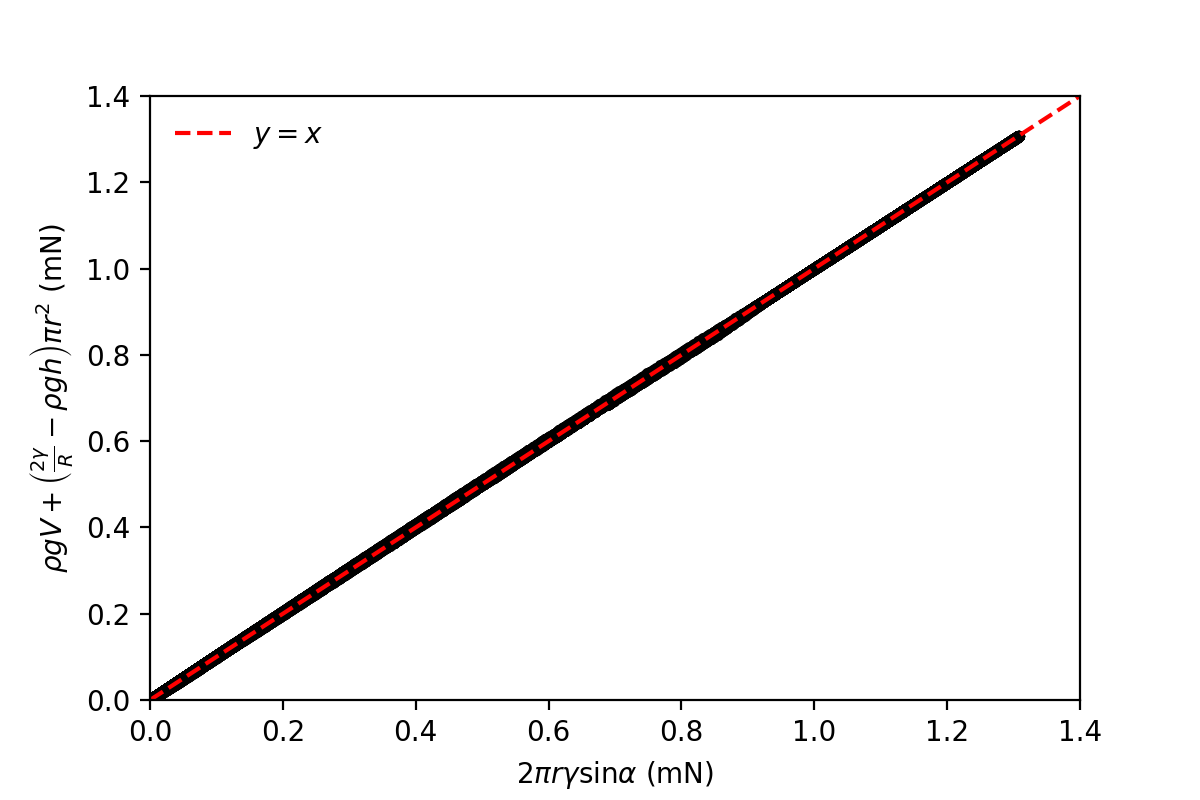} 
 \caption{Verification of the Furmidge-like relation}\label{fig:Furmidge}
\end{figure}

For each intermediate equilibrium profile with fixed patch radius $r:=y_{B}$, the bubble has a vapour contact angle $\alpha $ 
at the plate and, with $V$ the volume of the corresponding
intermediate bubble, by Archimedes principle, 
\begin{eqnarray} \label{Furm}
2\pi r\gamma \sin \alpha  &=&\rho gV+\left( p-p_{1}\right) \pi r^{2} \cr
  &=&\rho gV+\left( \frac{2\gamma }{R}-\rho
gh\right) \pi r^{2}
\end{eqnarray}
translating the vertical balance of surface versus gravitational
forces, in the spirit of Furmidge, \cite{Fato}. 
We have checked this equation numerically, see Fig.~\ref{fig:Furmidge}.

See \cite{Fato}, Eq. (12), where $g$ was changed to $-g$ while considering bubbles instead of drops.

This is in particular true when $V=V_{\max }$\ and $\alpha =a_{B}$, where $B$ reaches the terminal value $\sin \alpha $.

We give the profiles with $y_B=0.49$ mm for $R\in [0.25,3.68]$ mm with a step $\Delta R=0.002$ mm. 
Subsequently, we can compute numerically the volume of
the bubble, namely 
\begin{equation}
V=\pi \int_{0}^{h}y^{2}(x)dx.
\end{equation}
where $h$ is the height of the bubble.

\begin{figure}[htp!] 
 \centering
 \includegraphics[width=\textwidth]{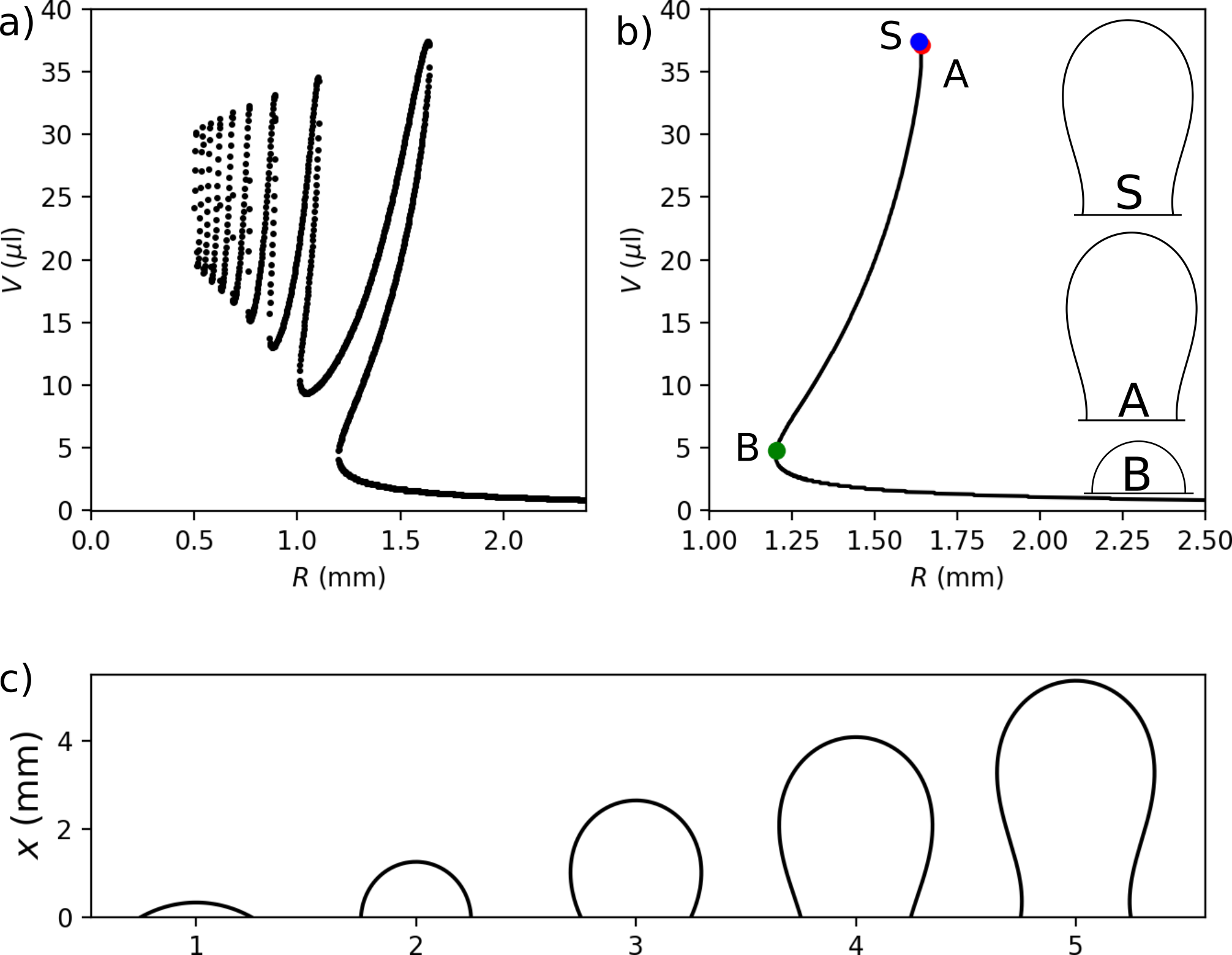} 
 \caption{a) a plot of $V$ against $R$; b) a plot of $V$ against $R$ till the maximum volume is reached at $S$; c) the corresponding sequence of bubbles}\label{fig:3}
\end{figure}

In Fig.~\ref{fig:3}a, we represent $V$ versus $R$ for $y_B=1.23$ mm. While following $R$ starting from large values, the volume of
the bubble is seen to increase before it reaches a maximum volume at $S$
just after the appearance of a first neck in its profile which appears at $B$
when the tangent at the basis of the bubble becomes vertical for the second
time after $A$; see Fig.~\ref{fig:3}b. So a neck appears if the bubble profile shows two vertical tangents or if there is a local minimum of $y$ other than 0 and $r$.

For $r > r_{crit}=2.198$ mm, there are configurations for which the neck does not appear in the equilibrium regime.

We get a sequence of bubbles for increasing volume from $V=0$ till $%
V=V_{\max }$ that can be considered as a succession of equilibrium states
and we are restricted to the curve shown in Fig.~\ref{fig:3}b.

Some of the profiles obtained are shown in Fig.~\ref{fig:3}c (the extreme right one corresponds to point $%
S$ with the maximum volume).

\begin{figure}[htp!] 
 \centering
  \includegraphics[width=\textwidth]{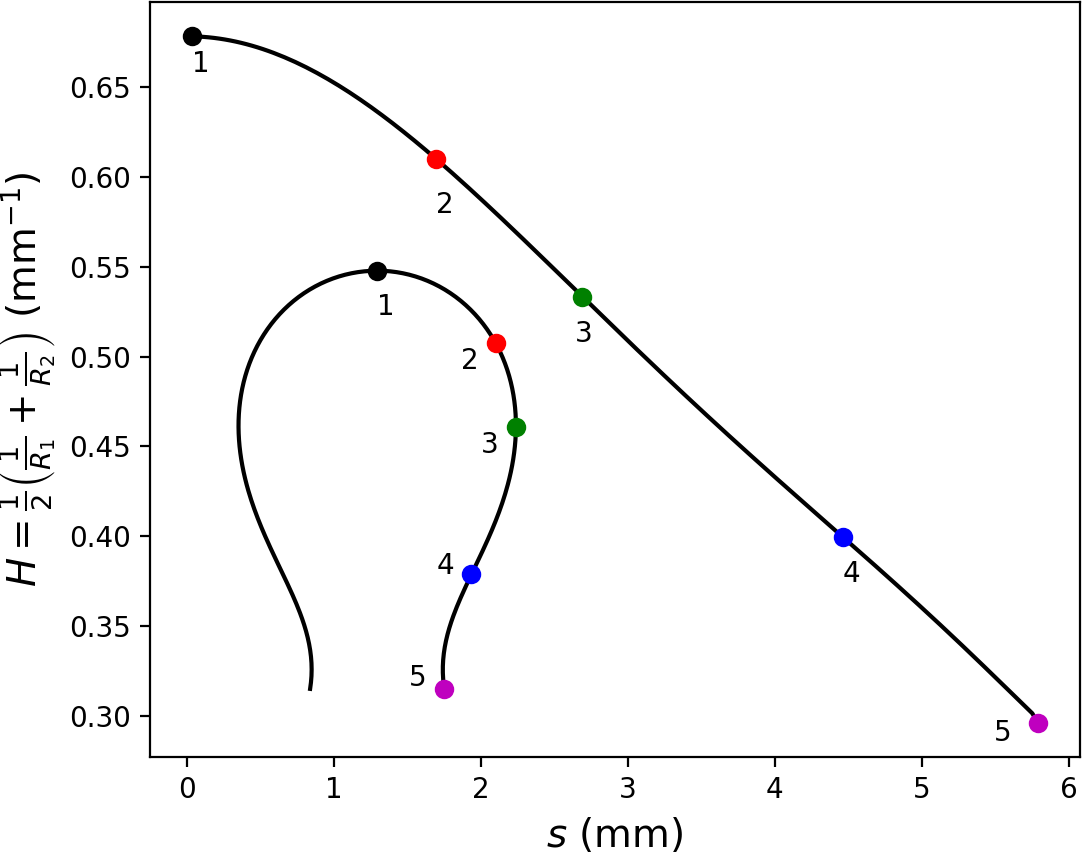} 
 \caption{$r = 0.76$ mm, $V_{\max}=22.79$ $\mu$l. Plot of the mean curvature along a terminal bubble profile ready for detachment.} 
\label{fig:H} 
\end{figure}

Fig. \ref{fig:H} shows the evolution of the mean curvature along the profile corresponding to such a 'terminal' equilibrium profile before detachment. It decreases while descending from the apex to the bottom of the bubble but remains positive.

Consider a bubble of an increasing volume $V$ with the contact line pinned
on a patch with radius $r=1.23$ mm. Beyond a volume
$V_{\max}$, no solution either stable or unstable of LH equations, \cite{LH91},
were found numerically. After reaching $V_{\max}$, we expect to enter a non-equilibrium regime, leading eventually to the detachment of the bubble.

The  height of the bubble at this volume is $h$. From the LH
expressions, \cite{LH91}, we are able to compute the equilibrium profiles and then, the
surface areas, contact angle, energy and pressure.

\textbf{Contact angle at the plate: }The contact angle $\alpha :=a_{B}$ at
the plate is computed from the tangent to the profile in contact with the
plate. We let $\theta =\pi -\alpha .$

\textbf{Energy}, \cite{Pitts74}: The different contributions to the total energy are 

- Gravitational and liquid/vapour interfacial energy: 
\begin{equation}
dE_{LV}=\left( 2\pi \gamma y\left( x\right) \frac{ds}{dx}-\rho g\pi y\left(
x\right) ^{2}(h-x)\right) dx
\end{equation}
\begin{equation}
E_{LV}=\int_{0}^{h}\frac{2\pi \gamma y\left( x\right) x}{\sin a\left(
x\right) }dx-\rho ghV+\rho g\pi \int_{0}^{h}y\left( x\right) ^{2}xdx
\end{equation}

- Surface energy at the Solid/Vapour and Solid/Liquid interfaces. With $\gamma
_{SV}-\gamma _{SL}=\gamma \cos \theta _{Y}$ (by Young-Dupr\'{e} equation
where $\theta _{Y}$ is the equilibrium Young angle of water),

\begin{equation}
E_{SV}+E_{SL}=\mbox{constant}+\pi (\gamma
_{SV}-\gamma _{SL}) r^{2}.
\end{equation}

The total mechanical energy (surface tension plus gravitational) $%
E=E_{LV}+E_{SV}+E_{SL}$ then reads, dropping the constant

\begin{equation} \label{Energy}
E=\pi (\gamma
_{SV}-\gamma _{SL}) r^{2}+\int_{0}^{h}\frac{2\pi \gamma y\left( x\right) x}{\sin
a\left( x\right) }dx-\rho ghV+\rho g\pi \int_{0}^{h}y\left( x\right) ^{2}xdx
\end{equation}
Due to the contact line pinning, $E_{SV}+E_{LV}$ does not change during the
evolution of the bubble.

\begin{figure}[htp!] 
 \centering
 \includegraphics[width=\textwidth]{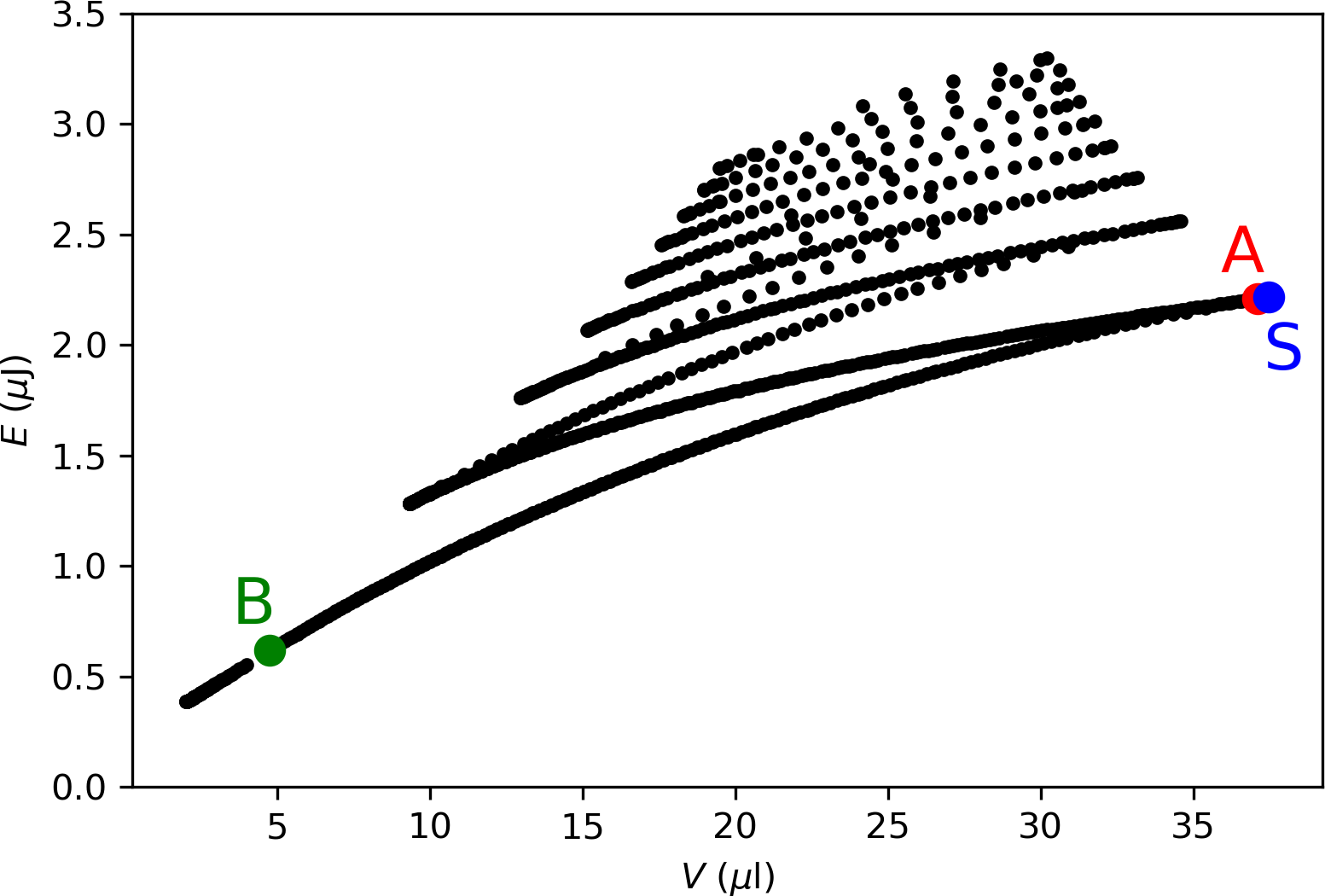} 
 \caption{Total mechanical energy for $y_B=1.23$ mm.} \label{fig:X}
\end{figure}

The total mechanical energy versus volume V is shown in Fig.~\ref{fig:X} where we observe that $E$ increases until a maximal value $%
V_{\max }$. After having reached $V_{\max }$, the energy decreases together
with $V$. The point $\left(
V_{\max ,}E_{\max }\right) $ is a singular point and the bubble will  evolve towards detachment when V increases beyond $V_{\max }.$ 

We have the following formula under the Dirichlet b.c.
\begin{equation}
    \lim_{V\rightarrow 0^+}E(V)=\gamma\pi r^2+(\gamma_{SV}-\gamma_{SL})\pi r^2
\end{equation}

For  Neumann b.c., we expect $\lim_{V\rightarrow 0^+}E(V)=0$. 

Fig. \ref{fig:X}, a plot of $E$ against $V$ shows that $E$ increases until a maximal volume $V_{\max}$ is reached. After having reached $V_{\max}$, the energy decreases together with $V$. The point $(V_{\max},E_{\max})$ is a singular point and the bubble will evolve towards detachment when $V>V_{\max}$.

\textbf{Pressure: }The pressure in the bubble is computed as

\begin{equation}
p=p_{1}+\frac{2\gamma }{R}-\rho gh
\end{equation}
where $p_{1}$ is the liquid pressure at the plate, taken as $p_{atm}+\rho gH$
where $H$ is the height of water above the plate assuming $H> h$ and $%
p_{atm}=10^{5}$ Pa is the atmospheric pressure$.$

If we are to apply this theory to a pinned bubble growing on a hydrophobic patch (Young angle $\theta_2$) immersed in a hydrophilic matrix (Young angle $\theta_1$), we need to impose
the condition $\theta_1<\pi- \alpha $ for all angles $\alpha$ appearing in the formation history of the bubble, in particular the terminal one $\alpha$ (when $V=V_{\max}$) but not only. This is possible and the worst
case where this condition could be violated is after point $A$ (where the tangent to the bubble is vertical for the first time). 
If this condition is not met, the bubble overflows the patch before reaching its terminal state for detachment.

The condition $\pi - \alpha < \theta_2$ for all angles $\alpha$ appearing in the formation history of the bubble is also required for pinning. 
If this condition is not met, the bubble would de-pin before reaching its terminal state for detachment.

It seems to suggest that growing a bubble in this way on a hydrophobic patch is impossible because then $\pi - \alpha >\theta_2$ at least at the early stages of the bubble  growth with small $\alpha$ and so large $\pi-\alpha$, unless ideally $\theta_2=\pi$. In fact (under the $\Delta T_{onb}$ conditions below), only on the hydrophobic patch do micro-bubbles grow, invade the patch with radius $r$ and then the initial condition for LH should be started with a macro-bubble making the angle $\theta_2$ on the liquid side.
 
\begin{figure}[htp!] 
 \centering
 \includegraphics[width=\textwidth]{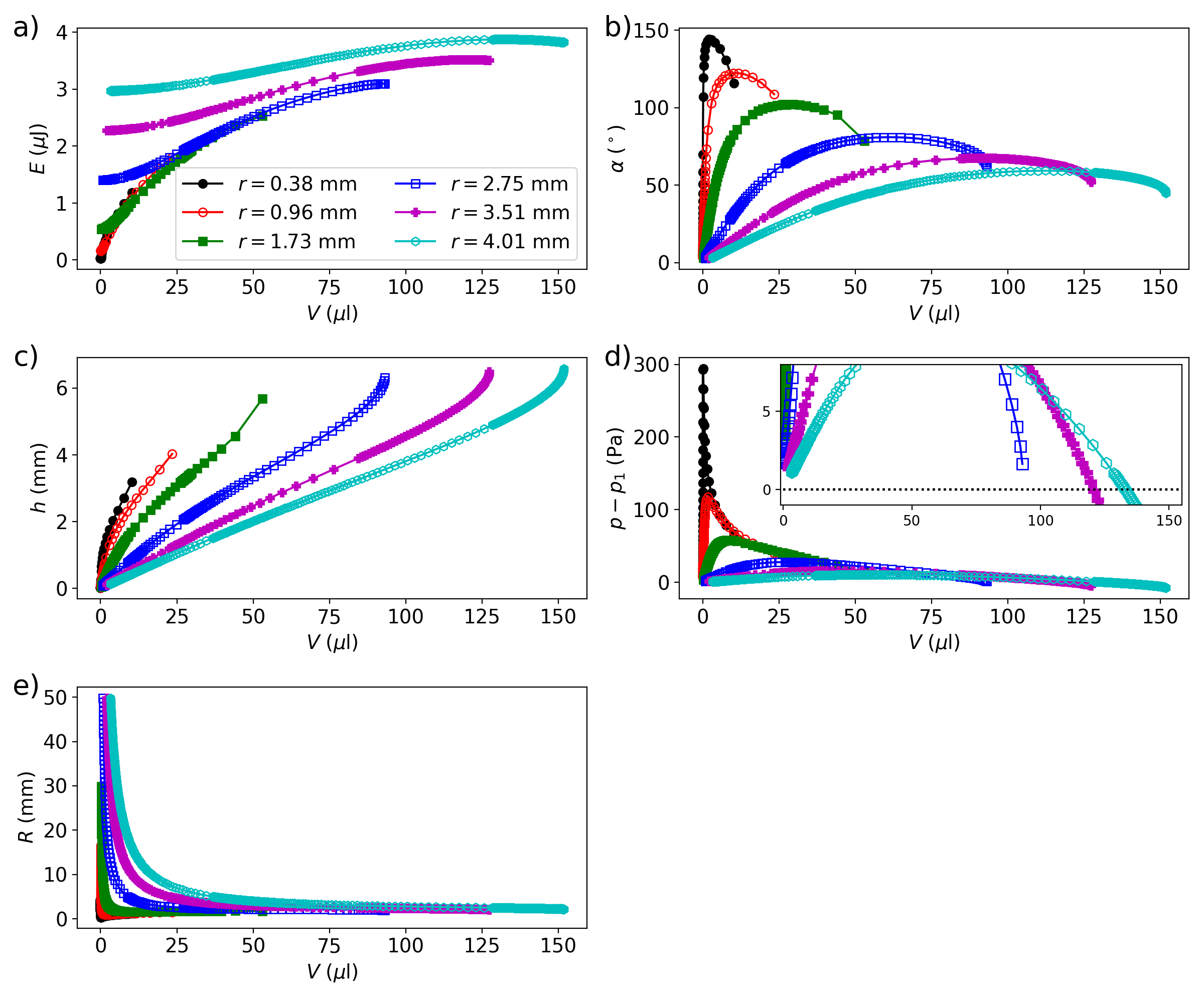} 
 \caption{$E$, $\alpha$, $h$, $p-p_1$ and $R$ versus $V$ under various Dirichlet b.c. The 'label' code of the radii displayed in a) also applies to b), c), d) e).} \label{fig:Dirichlet}
\end{figure} 

At fixed $r$ the variables $E$, $\alpha$, $h$, $p$ and $R$ can all derived as  functions of $V$ increasing till $V_{\max}$ as shown in Fig.~\ref{fig:Dirichlet}. The Furmidge identity Eq. \ref{Furm} gives a relation between $(E,\alpha,h, p, R)$.

\subsection{Bubble with prescribed contact angle $\alpha =a_{B}$ at the plate (Neumann boundary condition)}

With patches, one confines the bubbles to a contact area which is smaller than what they would have without pinning. Without patches, we now consider the growth of a micro-bubble on a pure substrate having Young water contact angle $\pi-\alpha=\theta_Y$. Here again, the plate is heated uniformly resulting in the formation of water vapour and so of bubble growth. The substrate is hydrophobic
(hydrophilic) if $\alpha<\pi/2$ ($\alpha>\pi/2$). 

In the range $\alpha\in[45^\circ,80^\circ]$, it appears that no neck exists till exiting the quasi-equilibrium regime.

\begin{figure}[htp!] 
 \centering
  \includegraphics[width=\textwidth]{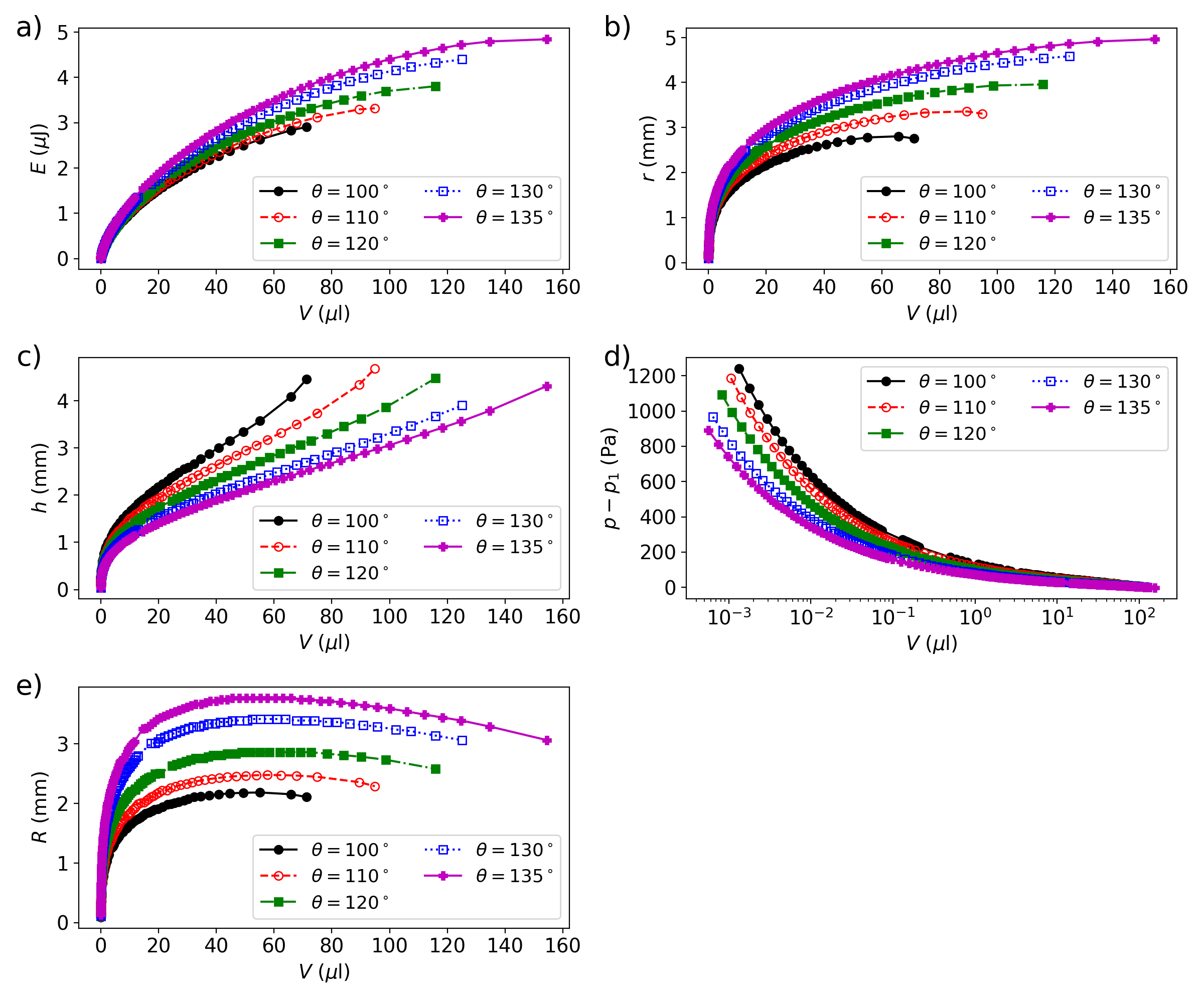} 
 \caption{$E$, $r$, $h$, $p-p_1$ and $R$ versus $V$ under various Neumann b.c.}\label{fig:Neumann}
\end{figure}

Fix $\alpha =a_{B}\in \left( 0,\pi \right) $ and integrate $\left( S\right) $
till this contact angle is met at the plate. In Fig.~\ref{fig:Neumann} we show the characteristics $E$, $r$, $h$, $p$ and $R$ of the bubble till it
reaches the maximal volume obtained when the energy is maximal. The
sequence of bubble shapes for $\alpha=60^\circ$ is shown in Fig.~\ref{fig:X3}. The radius at the
base $r=y_{B}$ of the bubble now varies with $R$, grows and eventually decreases with the volume $V$ till some $V_{\max}$ is reached corresponding to a maximal mechanical
energy of the bubble.

\begin{figure}[htp!] 
 \centering
 \includegraphics[width=\textwidth]{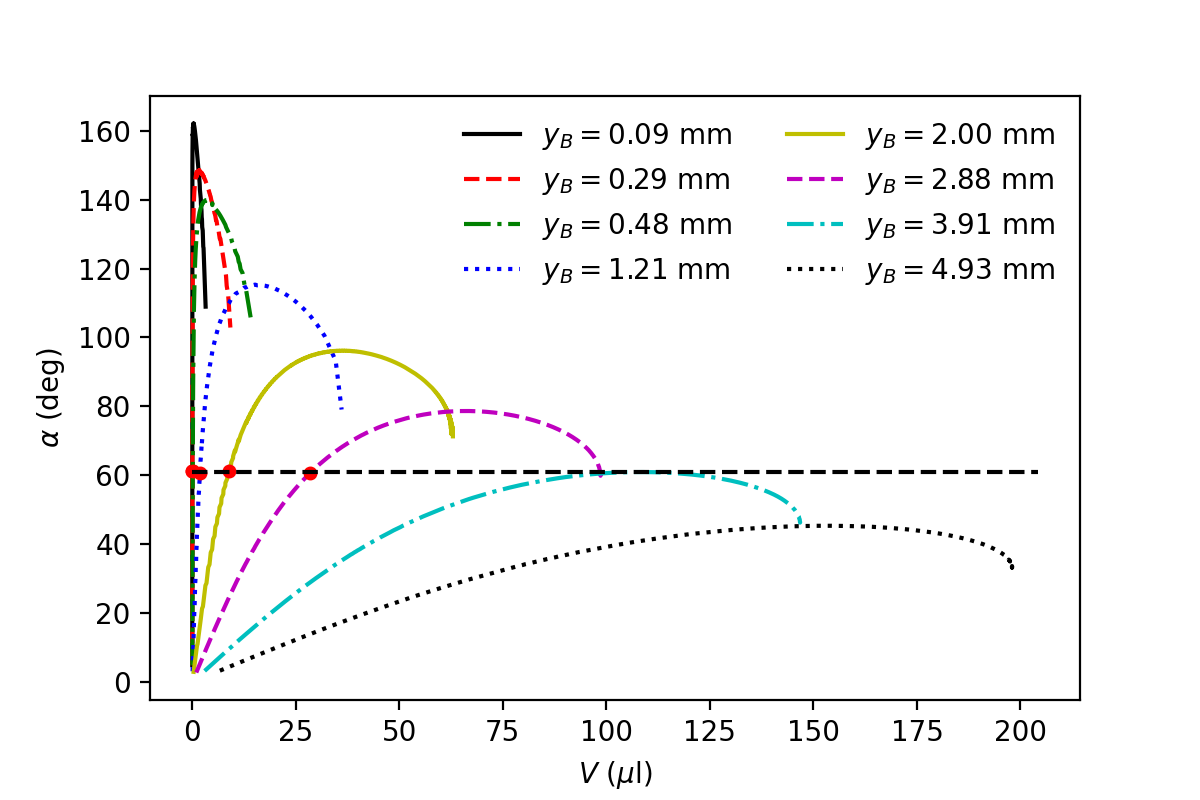} 
 \caption{Dependence of the angle $\alpha$ on the bubble volume $V$ for different bubble contact radii $r$=$y_B$.
 }\label{fig:X2}
\end{figure}

\begin{figure}[htp!] 
 \centering
 \includegraphics[width=\textwidth]{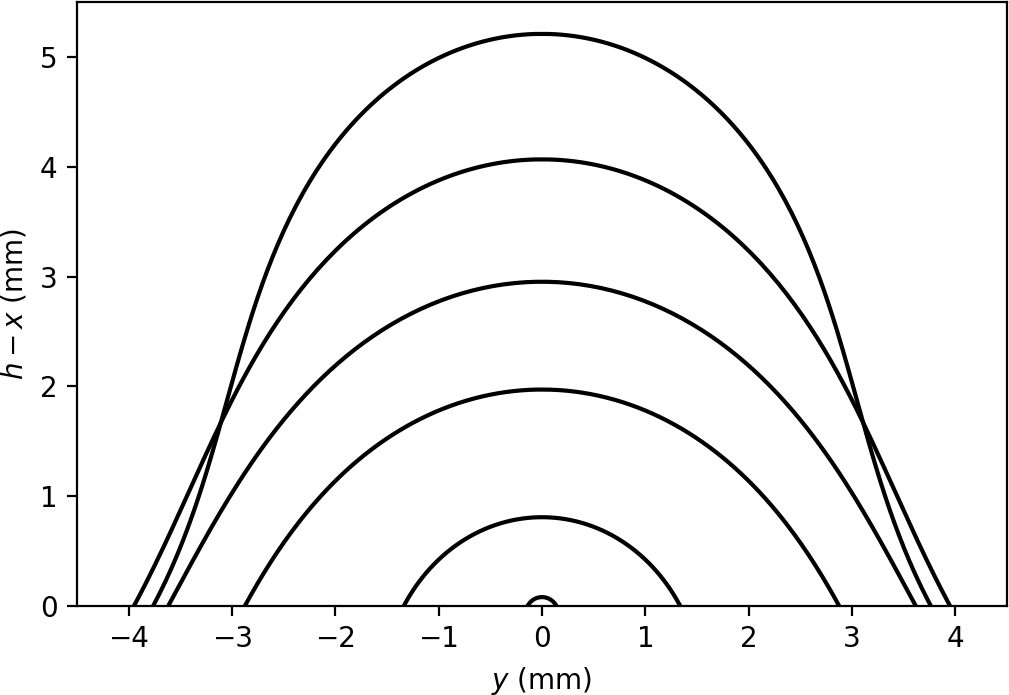} 
 \caption{Evolution of the drop profile computed for water with Neumann boundary condition $\alpha=60^\circ$. The drop radius first grows with the volume and then decreases. The highest profile, corresponding to $V_{\max}$, has a diameter about 7.5 mm.
 }
 \label{fig:X3}
\end{figure}

Consider the particular case $\alpha=60^\circ$ corresponding to a hydrophobic substrate with liquid contact angle $\theta=120^\circ$. We select a range of increasing values for the contact radius of the bubble $\{y_B\}$ and for each value we compute the volume $V$ and the contact angle $\alpha$ in the stable region until the maximum volume $V_{\max}$ is reached. Then, we represent $\alpha$ versus $V$ for each $y_B$ and we find the intersection (if it exists) with $\alpha=a_B=60^\circ$ as shown in Fig.~\ref{fig:X2}. From Fig.~\ref{fig:Neumann}b, there is a critical radius of the bubble $y_B^{c}=3.88$ mm for which it is not possible to find a stable solution with $\alpha=60^\circ$ for $y_B>y_B^c$. The fixed contact angle condition means that there is no contact line pinning. The contact radius will increase with the bubble volume if $\alpha>60^\circ$ or it will decrease if $\alpha<60^\circ$ until the condition $\alpha=60^\circ$ is satisfied. Starting from a small bubble volume, the contact radius will increase with the bubble volume to maintain the angle $\alpha$. After a critical volume is reached,  the bubble cannot sustain an angle  of $60^\circ$ for any contact radius and then, it will evolve and eventually detach from the substrate. Therefore, the Neumann condition implies the existence of a maximum volume $V_{\max}^{\alpha_c}$ defined as the largest volume for which we still have a quasi-static solution with an angle of $\alpha=60^\circ$. 

The evolution of the profile with a prescribed angle $\alpha=60^\circ$ until it reaches the maximum volume $V_{\max}$ is shown in Fig.~\ref{fig:X3}. This gives the contact radius of the bubble on a flat substrate when reaching instability, just before detachment, a problem addressed by many authors, see \cite{KGP21} and references therein.

In Fig. \ref{fig:X2}, for the values of $r$=$y_B$ such that the curves intersect  the dashed line corresponding to $\alpha=60^\circ$,  depinning when $\theta_1=120^\circ$  will occur. The smallest $y_B$ avoiding depinning before detachment is about 3.91 mm. If $\theta_1=65^\circ$, the dashed line should be set at $\alpha=115^\circ$, corresponding to a minimal $r=1.21$ mm. 

\subsection{Bubble profiles at zero gravity}

When $g=0$, the system $\left( S\right) $ is explicitly integrable, giving
the spherical-cap equation of the profile as ($\alpha =a_{B}$) 
\begin{equation}
x=R\left( 1-\cos a \right) ,\text{ }y=R\sin a \text{ and }%
s=R a 
\end{equation}
obeying 
\begin{equation}
y^{2}=R^{2}-\left( R-x\right) ^{2}. 
\end{equation}
A spherical cap with fixed contact radius $y_{B}=r$ at its bottom plane only
exists if $R\geq r$ and then it has height $h=x_{B}$ given by: 
\begin{equation}
h=R\mp \sqrt{R^{2}-r^{2}}\text{ } 
\end{equation}

The volume of the spherical cap is $V=\frac{\pi h}{6}(3r^{2}+h^{2}).$
Starting with a large $R$ and corresponding $h<R$, the volume steadily
increases with $R$ decreasing until $R=r$. After having reached $R=r$, a
branch with $h>R$ can be launched along which the volume increases with $R$
increasing and it has no upper bound. 

If we now consider spherical caps with fixed contact angle $\alpha =a_{B}$ at its bottom plane, $ V=\frac{\pi R^{3}}{3}(2+\cos \alpha )\left( 1-\cos
\alpha \right) ^{2}$ with  $r=y_{B}=R\sin a_{B}=R\sin \alpha $ growing linearly with $R$ starting from $0$. Note $R\geq r$ for all choices of $\alpha $ in $\left( 0,\pi \right)$. The volume grows like $R^{3}$ with
no upper bound either. Note that (with here
$p_1 = p_{atm}$)

\begin{equation}
h=R(1\mp \cos\alpha)\text{ and} 
\end{equation}

\begin{equation}
p-p_1=\frac{2\gamma}{R}-\rho g h=\frac{2\gamma}{R}-\rho g R (1\mp \cos\alpha)\text{ } 
\end{equation}

so that, for each fixed $r$, all physical bubble profile quantities of interest can be expressed in terms of $\alpha$, the vapour contact angle at the base of the bubble.

\subsection{Onset of boiling: micro-bubble nucleation}

In this Section, we assume that gravity is negligible in the first place (the approximation $g=0$).
Consider a radius $R$ spherical cap bubble resting on a flat substrate having contact radius $r$ and vapour contact angle $\alpha =a_{B}$. It has
height $h=R/\left( 1+\cos \alpha \right) .$ Its mechanical energy is

\begin{eqnarray}
E &=&E_{SV}+E_{SL}+E_{LV}\nonumber\\
E_{SV}+E_{SL}&=&\pi (\gamma
_{SV}-\gamma _{SL}) r^{2}=\pi \gamma r^{2}\cos \theta _{Y}\nonumber\\
E_{LV}&=&2\pi \gamma \int_{0}^{R\alpha }y\left(
s\right) ds=2\pi \gamma R^{2}\left( 1-\cos \alpha \right) = \frac{2\pi \gamma r^{2}}{1+\cos \alpha } 
\end{eqnarray}
where $\gamma
_{SV}-\gamma _{SL} =\gamma \cos \theta _{Y}$.

The angle $\pi -\alpha $\ is the receding water contact angle and if
hysteresis is neglected, $\alpha =\pi -\theta _{Y}$. We assume that the bubble obeys $pV=n\mathcal{R}T$ where $n$ is the number of vapour moles
and $T$ is uniform and constant in the vapour phase. Now, consistently with (14)(15) of \cite{PC09},
\begin{equation} \label{Vol}
    V=\frac{\pi r^3}{3}\frac{(2+\cos\alpha)(1-\cos\alpha)}{\sin\alpha(1+\cos\alpha)}=\frac{\pi }{3}R^{3}\left( \cos ^{3}\alpha -3\cos \alpha
+2\right)
\end{equation}
\begin{equation}
E_{\Delta T}=-A n\Delta T=-A\Delta T{\frac{pV}{\mathcal{R}T}} 
\end{equation}
is the free energy lost upon vapourization of the volume $V$ at a
temperature $T=373+\Delta T$ above the coexistence temperature. Here $A=157$
J/(mol.K) is related to the latent heat of water at a temperature close
to boiling, see Eq. \ref{AA}.
The free energy of vapourization into a bubble reads 
\begin{equation}
F=E_{SV}+E_{SL}+E_{LV}+E_{\Delta T}=r^{2}f_{1}-r^{3}f_{2}\Delta T
\end{equation}
where $f_{1}$ and $f_{2}$ as functions of $\alpha $ are:

\begin{eqnarray}
f_{1}\left( \alpha \right) &=&\pi \gamma \left( -\cos \alpha+\frac{2}{%
1+\cos \alpha }\right) \nonumber \\
f_{2}\left( \alpha \right) &=&\frac{\pi A p}{\mathcal{R}T}{\frac{(1-\cos
\alpha )(2+\cos \alpha )}{\sin \alpha (1+\cos \alpha )}}
\end{eqnarray}
where one can choose $\alpha = \pi-\theta _{Y}$. The free energy tends to
decrease because the energy per mole of vapour is lower than the energy per mole of
the liquid but it tends to increase because an interface between the liquid
and the vapour is created. Since the number of molecules that have changed from
liquid to vapour varies as the cube of the contact radius $r$ of the bubble,
whereas the area of the interface goes as $r^{2}$, the free energy at first
increases with increasing contact radius before reaching a maximum value at
a critical radius $r^{*}.$ The free energy thus has a maximum $F^{*}=F(r^{*})$ at
the critical radius $r^{*}$ given by 
\begin{eqnarray} \label{star}
r^{*}=\frac{2f_{1}}{3f_{2}\Delta T}={\cal O}\left( \alpha /\Delta T\right)\\ 
F^{*}=\frac{f_{1}}{3}r^{*2}=\frac{4f_{1}^{3}}{27f_{2}^{2}\left( \Delta
T\right) ^{2}}={\cal O}\left( \alpha ^{4}/\left( \Delta T\right) ^{2}\right)
\end{eqnarray}

The radius $r^{*}$ is the critical
value of the footprint radius of the bubble below which it cannot grow on a
free unpatterned substrate, (see \cite{LF67}, problem 1, page 570). The energy barrier is effective only if higher than thermal
or disorder fluctuations or some activation energy, say $F_{0}$. The onset
of boiling is then defined by $\Delta T_{onb}$ such that $F^{*}=F_{0}$%
, namely from Eq. ~\ref{star},
\begin{equation}\label{eq:Tonb}
\Delta T_{onb}=\left( \frac{4f_{1}^{3}}{27f_{2}^{2}F_{0}}\right)
^{1/2} 
\end{equation}

\begin{figure}[htp!] 
 \centering
 \includegraphics[width=\textwidth]{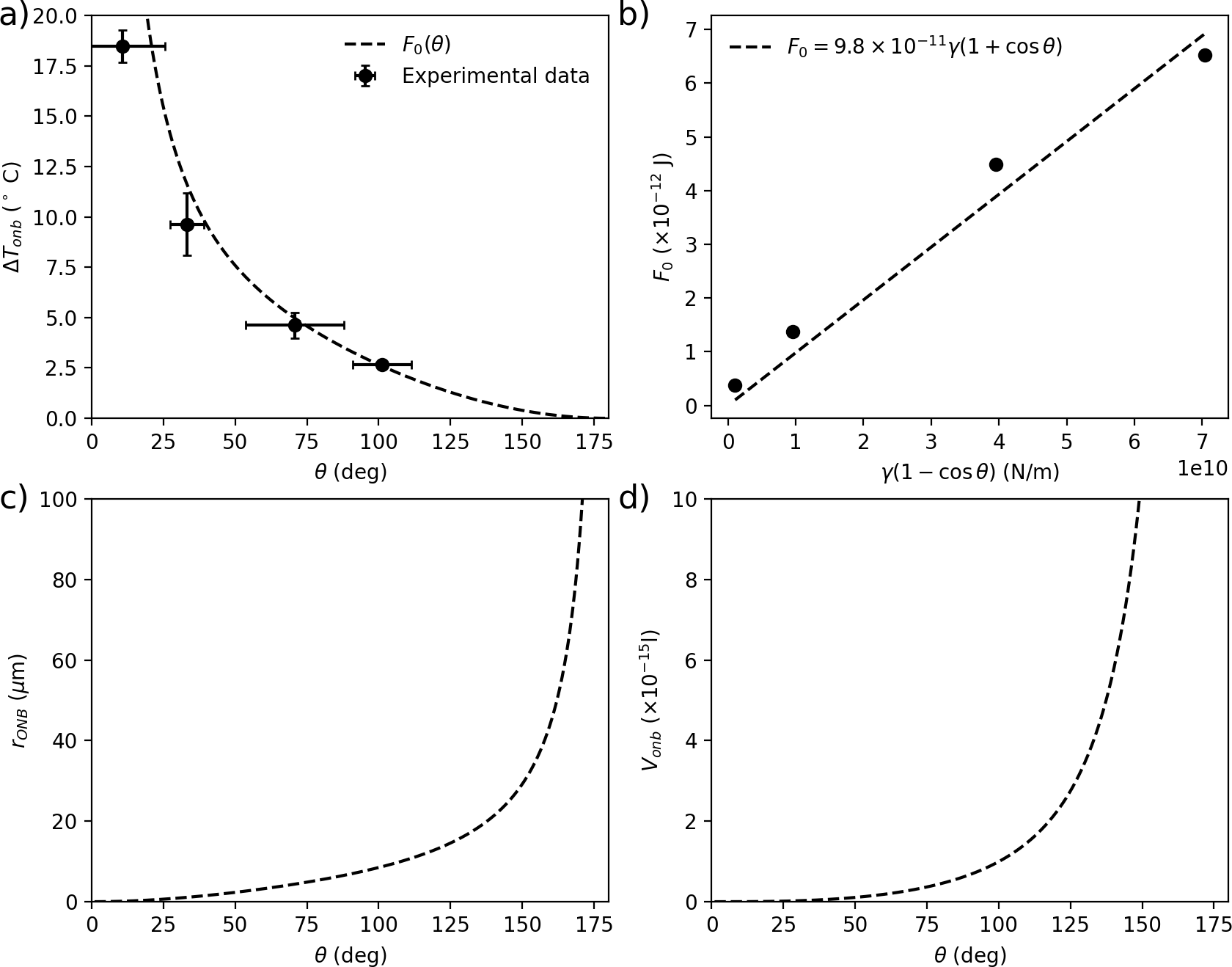} 
 \caption{$\Delta T_{onb}$, activation energy, $r_{onb}$ and $V_{onb}$ against the liquid Young contact angle $\theta$=$\theta_Y$.}\label{fig:Tonb1}
\end{figure}

We can expect a dependence of the activation energy $F_0$ with the substrate wettability, or more precisely, with the work of adhesion $W=\gamma(1+\cos\theta_Y)$. We can then compute the activation energy from Eq.~\ref{eq:Tonb}  to fit the experimental data of Fig.~3B in Ref. \cite{Bourdon15}. In Fig.~\ref{fig:Tonb1}b we observe a linear dependence between the computed activation energy and the work of adhesion that can be fitted with a simple straight line. Then, we introduce the fitted parameters in Eq.~\ref{eq:Tonb}.

We obtain $\Delta T_{onb}$ for various
values of $\theta _{Y}$,  as shown in Fig.~\ref{fig:Tonb1}a. 
The value $r_{onb}$ in Fig.~\ref{fig:Tonb1}c is obtained while plugging $\Delta T=\Delta T_{\text{onb}}$ in the expression of $r^{*}$ in  Eq. ~\ref{star}. We get 
\begin{equation} \label{ronb}
    r_{onb}=\left( \frac{3F_{0}}{f_{1} }\right) ^{1/2}
\end{equation}
Similarly, the volume $V_{onb}$ on Fig.~\ref{fig:Tonb1}d of the bubble at the onset of boiling is obtained while plugging $r=r_{\text{onb}}$ in Eq. ~\ref{Vol}.

As shown in Fig.~\ref{fig:Tonb1}c, in a realistic range of $\theta$=$\theta_Y$,  a bubble of radius of order $10$
micrometers  is formed and grows and we can deal with the second stage,
detachment by gravity around $r=1$ mm, helped or not by pinning the bubble
at $r$ of order mm to maximize the flow. For such micro-droplets, the
approximation $g=0$ is justified a posteriori.

From the plot $\Delta T_{onb}$
against $\theta$=$\theta_Y$ (Fig.~\ref{fig:Tonb1}a), we see that the barrier is smaller in the hydrophobic than in the hydrophilic side, so that it is easier to grow micro-bubbles on the hydrophobic side so long as $\Delta T_{onb}%
<\Delta T$ and $\Delta T$ not too large so as not to activate the micro-bubbles on the hydrophilic side. 
For example for $\Delta T = 5 K$, micro-bubbles can grow on a substrate with liquid contact angle $\theta_2 >70^\circ$ but not on a substrate with liquid contact angle $\theta_1 <70^\circ$.


Micro-bubbles on the hydrophobic side, having a liquid contact angle $\theta_2 >90^\circ$ grow, coalesce and eventually form a macro-bubble with contact angle $\theta_2$ and sitting on the patch of radius $r$, which serves as the initial condition of subsequent bubble growth possibly leading to its detachment.

This analysis leads
to $T_{onb}\sim \,$ a few mK for a patch of order mm and $T_{%
onb}\sim $ a few K for a patch of order ten $\mu $m, or impurities
of corresponding sizes. One might start say at 105$^{\circ }$C to initiate
nucleation and then operate cycles at 100.1$^{\circ }$C. The model is for an
ideally structured substrate. The claim is that IF a bubble is formed then
it will grow and detach if $B-A\,(T-373)<0$. At $g=0$ and pressure close to
atmospheric pressure, the maximum of $B(V)$ is 268 J/mol for patch radius $%
r=15\,\mu $m and 29.4 J/mol for $r=0.15\,$mm and 2.97 J/mol for $r=1.5\,$mm.

We define the onset of boiling on the patterned surface as the smallest
temperature above which the bubble grows and reaches instability.

\section{Dynamics in quasi-static regime}
The bubble obeys $pV=n\mathcal{R}T$ where $n$ is the number of vapour moles
and $T$ is assumed uniform and constant in time in the vapour and on the
solid substrate, e.g. $T=(273+105)$ K. The vapour density is neglected compared
to the liquid density and $p$ is the thermodynamic pressure, uniform in the
bubble although not constant in time. The initial condition is taken as a Young-Laplace equilibrium bubble
covering a hydrophobic patch, with contact angle equal to the supplement of
the receding contact angle of water on the hydrophobic material. This initial condition  corresponds to either the time when the bubble first covers the patch, or the time shortly after detachment of a bubble leaving behind a minimal bubble. The pressure $p_{1}$ at the solid-liquid interface will remain constant,
typically a little above atmospheric pressure. We discuss here the slow
growth of the bubble through Young-Laplace equilibrium states, until
reaching instability. What occurs between successive bubbles is out of reach
here, it may take a small fraction of time so that neglecting it may lead to
a small error. Indeed on the one hand the unstable part of evolution is much
faster than the quasi-stable part. On the other hand, if the patch is not
yet covered by a bubble, heat transfer is more effective and therefore
faster.

At time $0\le t\le t_{\mathrm{unstable}}$, let us consider what
occurs between $t$ and $t+dt$ . Evaporation of one mole is
associated with an increase of enthalpy equal to the latent heat $ %
L=40.8\,$ kJ/mol and an increase of entropy $S=L/T$\ at
coexistence, $T=373\,$ K, so that the Gibbs free energy remains
constant. At a temperature slightly above coexistence, neglecting the
variation of $L$\ and $S$, the variation of Gibbs free energy will be, Ref. \cite{wiki} 
\begin{equation} \label{LG}
\Delta G=L-TS=-(T-373)S=-(T-373)L/373
\end{equation}

Taking into account the temperature dependence of $L$ from Ref.~\cite{Dehai20}
\begin{equation} \label{LL}
L(T)-L(373)\simeq (T-373){\frac{L(400)-L(300)}{100}}\simeq -40(T-373)\ 
\mathrm{J/mol} 
\end{equation}

and the temperature dependence of the vapour entropy,
\begin{equation} \label{DS}
\frac{dS}{dT}=\frac{\mathcal{R}}{T}\Rightarrow T\Delta S\simeq \mathcal{R}%
(T-373) 
\end{equation}
and neglecting that of the liquid leads for evaporation of $n$ moles to 
\begin{equation} \label{AA}
\Delta G=-n A(T-373) \ {\rm with} \  A\simeq 157
\ {\rm J/ (mol.K)}
\end{equation}

Evaporation presumably occurs at the triple line, but evaporated molecules
merge instantly with the rest of the bubble. Then the bubble, which
previously was in a Young-Laplace equilibrium shape, adjusts itself to a new
Young-Laplace equlibrium shape. If $dn$ moles were evaporated, the change in
gravitational and interfacial energies will be $dE$, given by integrals,
summing to $B\,dn$, resulting altogether in $dn\,(B-A\,(T-373))$ with $%
B=dE/\,dn$. Linear response ``\`{a} la Onsager'', \cite{LF67}, for a sequence of
quasi-equilibrium states then suggests 
\begin{equation} \label{dndt}
 dn=-2\pi r\lambda \,(B-A\,(T-373))\,dt  
\end{equation}

positive or negative, where $r$ is the patch radius and $\lambda $ is
the inverse of a time constant whose value is not predicted by
theory.

One can find the resulting function $V(t)$, the bubble volume $%
V_{\max }$ at $t_{\mathrm{unstable}}$ and the time to
unstability as function of patch radius. The vapour volume flux from a
surface of area $S_0$ and $S_0/(4r^{2})$ patches in this
approximate theory is $S_0V_{\max }/(4r^{2}t_{\mathrm{unstable}})$. The
factor $4$ is arbitrary. This flux could be maximised as function of
patch radius, and discussed as function of $T$.
Our goal could be to maximize the flow of evaporated vapour, proportional to
\begin{equation}
\frac{V_{\max }(r)}{r^{2}\Delta t(r)\,\lambda } 
\end{equation}

where $\Delta t(r)$ is the period, associated with the evolution of one
bubble, obtained from the slow quasi-equilibrium bubble evolution only. This
factor will decrease the optimal $r$, perhaps down to the order of 10$\mu $m
where $B$ comes into play.
The computation of $B$ starts with 
\begin{equation} \label{VB}
\frac{dV}{dn}=\frac{\mathcal{R}T}{p+Vdp/dV}, \qquad B=\frac{dE}{dn}=\frac{dE%
}{dV}\frac{dV}{dn} 
\end{equation}
where $dp/dV$ is a purely mechanical total derivative and $E$ is given by  Eq.\ref{Energy}.
Equilibrium shapes can be parametrized by the volume $V$, following
\cite{LH91}, yielding $dp/dV$ and $dE/dV$.

We can see that the effect of gravity on $B_{\max}$, and therefore on $B_{\max}/A$, is negligible for contact radius $r\le 1$ mm. From Fig. \ref{fig:B0}b, when $r$ is of order $1$ mm, $B<B_{\max }= 5$.

For such patch radii, considering the dynamics Eq. \ref{dndt} 
for $T=T_{\text{onb}} = (373+5)$ K, recalling $A\simeq 157$ J/(mol.K)%
, we conclude that $B\ll A\Delta T_{\text{onb}}$ so that, when gravity is
present (and also when gravity is absent), to a good approximation 
\begin{equation}
\frac{dn}{dt}=2\pi r\lambda \,A\,(T_{\text{onb}}-373))
\end{equation}

Hence

\begin{equation}
\frac{dn}{dt}=2\pi r\lambda \,A\,(T_{\text{onb}}-373))\,=\frac{1}{\mathcal{R}%
T_{\text{onb}}}\frac{d\left( pV\right) }{dt} 
\end{equation}

\begin{equation}
=\mathcal{R}T_{\text{onb}}\left[ \frac{Vdp\left( V\right) }{dt}+p\left(
V\right) \frac{dV}{dt}\right] =\frac{1}{\mathcal{R}T_{\text{onb}}}\frac{dV}{%
dt}\left[ p\left( V\right) +\frac{Vdp\left( V\right) }{dV}\right] 
\end{equation}

Therefore 
\begin{equation} \label{dVdt}
\frac{dV}{dt}=\frac{2\pi r\lambda \,A\,\mathcal{R}T_{\text{onb}}(T_{\text{onb%
}}-373))}{p\left( V\right) +Vp^{\prime }\left( V\right) },\, 
\end{equation}

giving, upon integration, a time evolution of $V$. 

When gravity is present, this integration must be done numerically using LH, as discussed now.

\subsection{Dynamics with gravity}

\begin{figure}[htp!] 
 \centering
 \includegraphics[width=\textwidth]{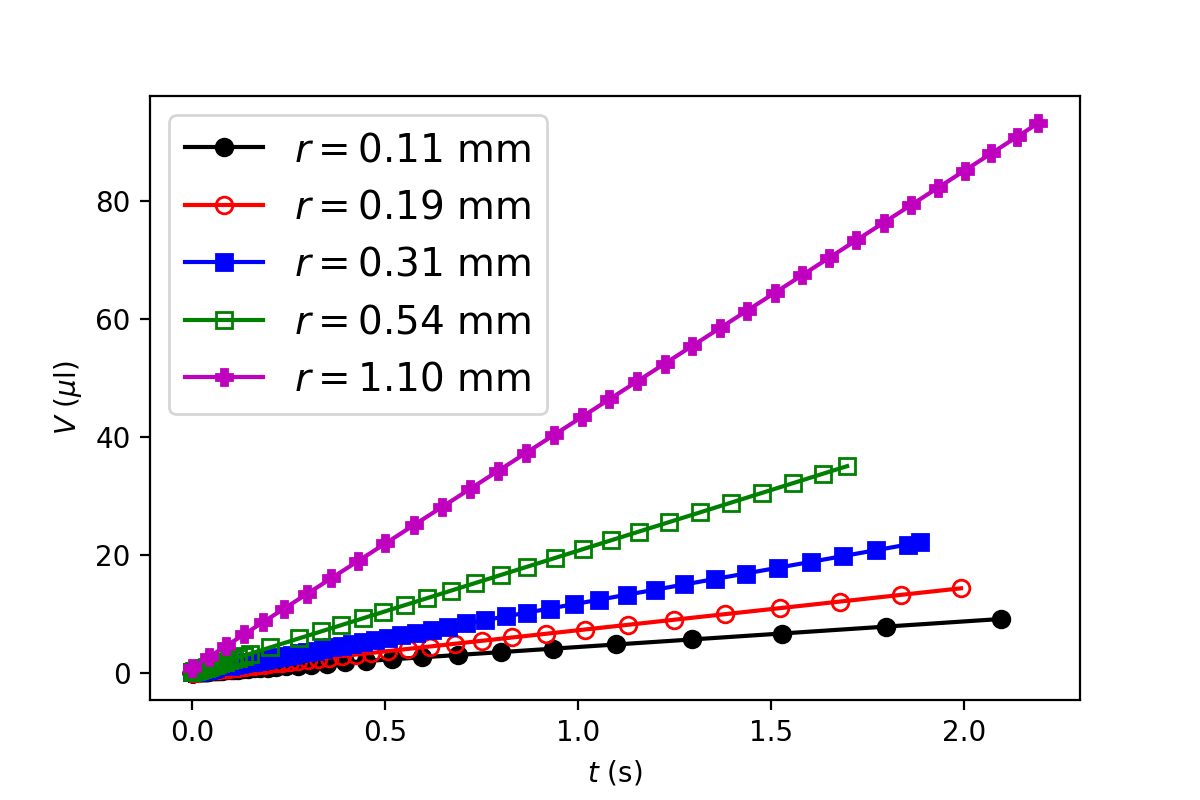} 
 \caption{Volume of the bubble versus time, as a solution to Eq. \ref{dVdt}, for various $r$.}\label{fig:Vvst}
\end{figure}

\begin{figure}[htp!] 
 \centering
 \includegraphics[width=\textwidth]{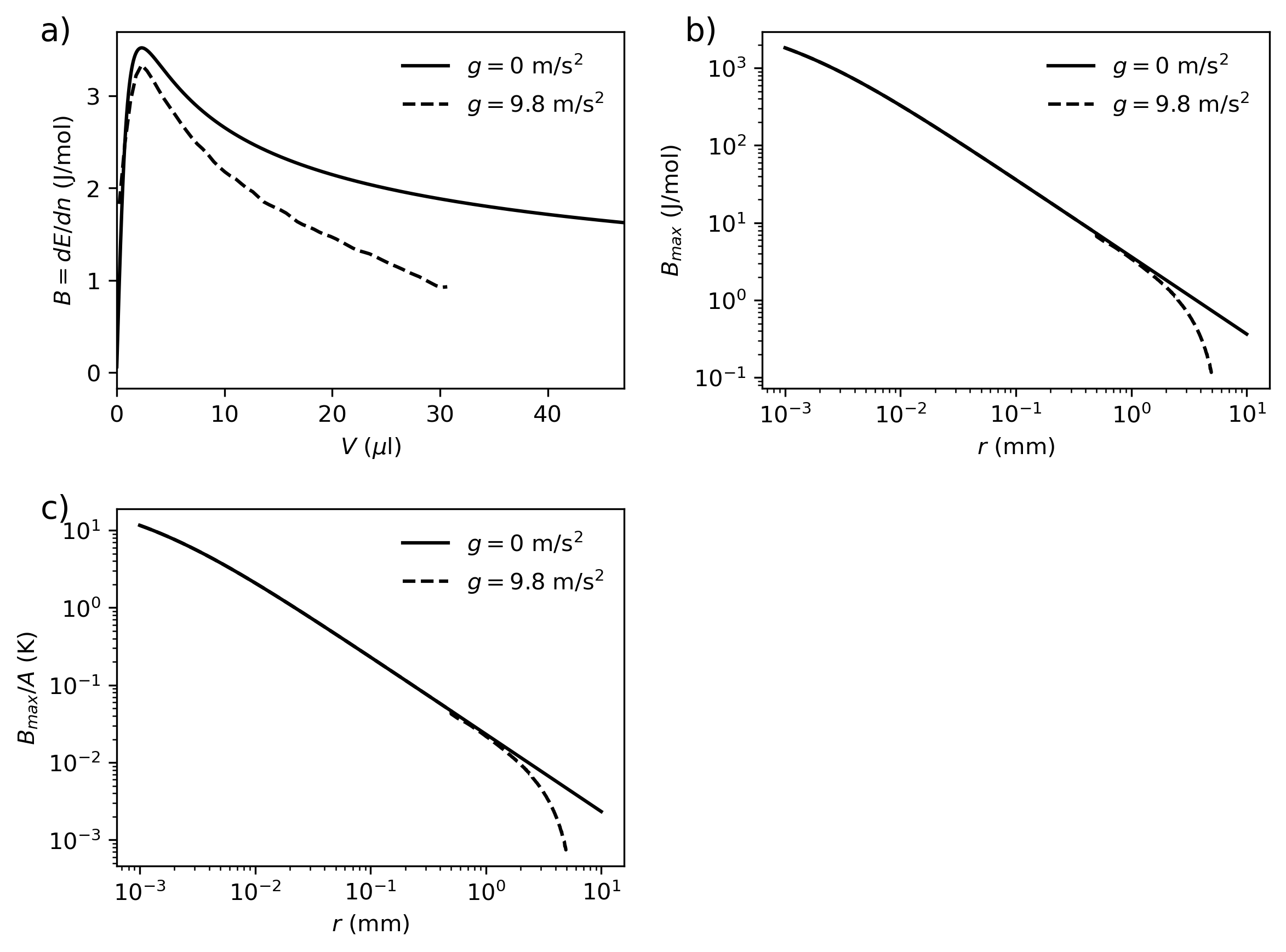} 
 \caption{ a) $B=dE/dn$ computed for $r=1$ mm. b) $B_{\max}$ obtained for the different contact radii. c) $B_{\max}/A$; with and without gravity.}\label{fig:B0}
\end{figure}

When gravity is present, Eq.~\ref{VB} must be solved numerically from the  profiles to compute the values of $B$ as shown in Fig.~\ref{fig:B0} when $g=9.8$.
In Fig.~\ref{fig:Vvst}, we plot $V$ against $t$ for various $r$ where we choose $\Delta T_{onb}=5$ K and  arbitrarily the value $\lambda=10^{7}  ({\rm s.J.m})^{-1} $. We observe a quasi-linear dependence of the volume with respect to time and this results from the fact that the relative variation of pressure against volume is negligible.

Once we have the time
evolution of $V$, the temporal evolution of the bubble profile
characteristics $\left( p,\alpha ,E,R,h\right) $ follows.

\subsection{Dynamics at zero gravity}

In this case, one can find an explicitly soluble  dynamical system giving the temporal evolution of the contact angle $\alpha$, up to the time of depinning where $%
\pi -\alpha =\theta _{1}^{\mathrm{rec}}$. Indeed, depinning occurs by the contact line moving out of the patch into the hydrophilic matrix. The final result is Eq. \ref{dalpdt}.

We have: 
\begin{equation}
R=\frac{2\gamma }{p-p_{1}},\text{ }\alpha =\arcsin (r/R), 
\end{equation}
\begin{eqnarray*}
E_{SV}+E_{SL} &=&\pi \gamma r^{2}\cos \theta _{Y}\text{, } \\
E_{LV} &=&\frac{2\pi \gamma r^{2}}{1+\cos \alpha }=2\pi \gamma
\,R^{2}(1-\cos \alpha ),
\end{eqnarray*}
\begin{equation}
E=E_{SV}+E_{SL}+E_{LV}=\pi \gamma r^{2}\cos \theta _{Y}+2\pi \gamma
R^{2}(1-\cos \alpha ) 
\end{equation}

\begin{equation}
V=\frac{\pi R^{3}}{3}(1-\cos \alpha )^{2}(2+\cos \alpha ) 
\end{equation}

At some contact angle away from $0$ and $\pi $, say $\pi /2$, consider a
variation $d\alpha $, then 
\begin{equation}
R=r\sin \alpha \ \Rightarrow \ dR\sim d\alpha ,\text{ }p-p_{1}=\frac{2\gamma 
}{R}\ \Rightarrow dp\sim d\alpha ,\text{ }V\ \text{as\ above}\Rightarrow \
dV\sim d\alpha , 
\end{equation}
\begin{equation}
pV=n\mathcal{R}T\ \Rightarrow \ dn\sim d\alpha ,\text{ }E\ \text{as\ above}%
\Rightarrow \ dE\sim d\alpha ,\text{ }\dots \ \Rightarrow B=O(1) 
\end{equation}
Therefore if $T-373$ is small, evaporation will stop before reaching this
angle, whereas if $T-373$ is large evaporation will continue beyond this
angle. How large and small will depend in particular upon $r$. At $g>0$ this
supports existence of $T_{onb}$.

More explicitly, for $g=0$, we have 
\begin{equation}
R(\alpha )=\frac{r}{\sin \alpha },\text{ }R^{\prime }(\alpha )=-\frac{r\cos
\alpha }{\sin ^{2}\alpha } 
\end{equation}
\begin{equation}
p(\alpha )=p_{1}+\frac{2\gamma }{R(\alpha )},\text{ }p^{\prime }(\alpha )=-%
\frac{2\gamma R^{\prime }(\alpha )}{R(\alpha )^{2}} 
\end{equation}
\begin{equation}
V(\alpha )=\frac{\pi R(\alpha )^{3}}{3}(1-\cos \alpha )^{2}(2+\cos \alpha ) 
\end{equation}
\begin{eqnarray*}
V^{\prime }(\alpha )&=&\pi R^{\prime }(\alpha )R(\alpha )^{2}(1-\cos \alpha
)^{2}(2+\cos \alpha )+\frac{\pi R(\alpha )^{3}}{3}\sin \alpha (1-\cos \alpha
)(1+2\cos \alpha ) \\
&=&\frac{\pi r^3}{4\cos^4\left(\frac{\alpha}{2}\right) }
\end{eqnarray*}

\begin{equation}
E(\alpha )=2\pi \gamma R(\alpha )^{2}(1-\cos \alpha )
\end{equation}
\begin{equation}
E^{\prime
}(\alpha )=4\pi \gamma R^{\prime }(\alpha )R(\alpha )(1-\cos \alpha )+2\pi
\gamma R(\alpha )^{2}\sin \alpha =8\pi \gamma r^2\frac{ \sin^4\left(\frac{\alpha}{2}\right)}{\sin^3\alpha }
\end{equation}

Therefore 
\begin{equation}\label{eq:B}
 \frac{dE}{dp}(\alpha )=\frac{E^{\prime }(\alpha )}{p^{\prime }(\alpha )},%
\text{ }\frac{dp}{dn}(\alpha )=\frac{\mathcal{R}T}{V(\alpha )+p(\alpha
)V^{\prime }(\alpha )/p^{\prime }(\alpha )},\text{ }B(\alpha )=\frac{dE}{dp}%
(\alpha )\,\frac{dp}{dn}(\alpha )   
\end{equation}

At this step, we have an explicit expression of both $B$ and $V$ against $\alpha$. 
Figure ~\ref{fig:B0}a shows the dependence of $B$ with the volume for $g=0$, $T=105^\circ$ C and $r=1$ mm where we set the pressure $p_1$ to the atmospheric pressure. The maximum value for $B$ obtained for any contact radius $r$ is shown in Fig.~\ref{fig:B0}b. Figure~\ref{fig:B0}c shows a plot of 
$B_{\max}/A$
as a function of $r$. From this, $B_{\max}/A\approx 0.01$ K for $r=1$ mm and $B_{\max}/A\approx 1$ K for $r = 10$ $\mu$m. 

\
Observing 

\begin{equation}
n^{\prime }(\alpha )=p^{\prime }(\alpha )\frac{dn}{dp}(\alpha ),
\end{equation}
we finally get the dynamical system on $\alpha$
\begin{equation} \label{dalpdt}
\frac{d\alpha }{dt}=\frac{dn}{dt}\frac{1}{n^{\prime }(\alpha )}=\frac{2\pi
\,r\lambda }{n^{\prime }(\alpha )}(A\,\mathcal{R}(T-373)-B(\alpha
))=:f(\alpha ) 
\end{equation}

Solving the autonomous equation $d\alpha /dt=f(\alpha )$ yields $\alpha (t)$%
, whence $R(t)=R\left( \alpha (t)\right)$,  $p(t)=p\left( \alpha (t)\right)
, V(t)=V\left( \alpha (t)\right) $. Knowing the values of each one of these variables versus $\alpha$, one could use $V$ or $p$ or $n=pV/(%
\mathcal{R}T)$ in place of $\alpha $, but the analog of $f$ would be defined
implicitly.




\section{Discussion}
\subsection{Effect of gravity}

\begin{figure}[htp!] 
 \centering
 \includegraphics[width=\textwidth]{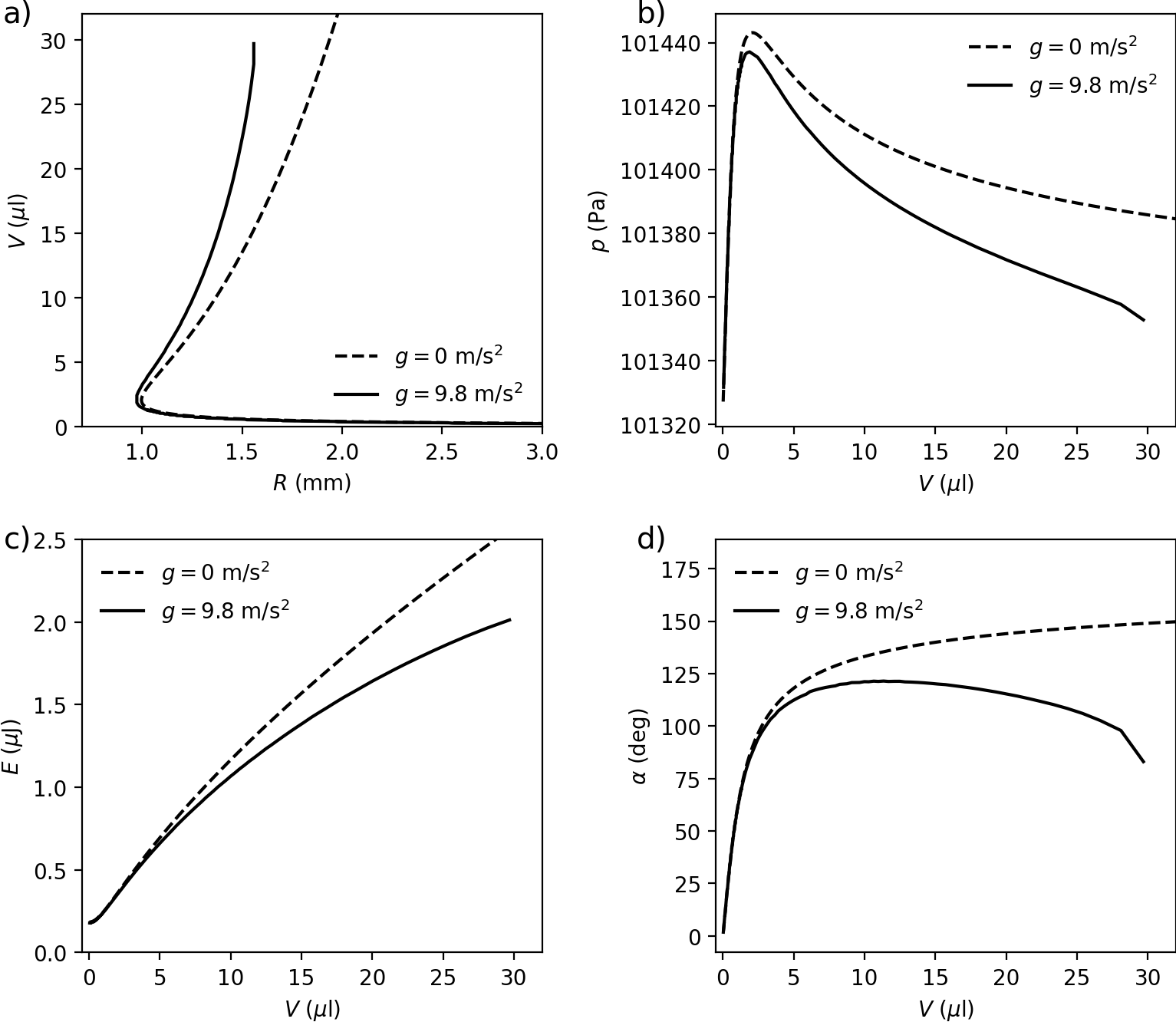} 
 \caption{a): $V$ versus $R$ and b), c), d): $p$, $E$, $\alpha$ versus $V$, for $g=0$ and $g=9.8$. Patch radius $r=1$mm.}\label{fig:6}
\end{figure}

Consider the particular case of a bubble with a fixed contact radius $r=1.0$ mm (Dirichlet boundary condition) at temperature $T=105 ^\circ$C, water surface tension $\gamma=0.0578$ N/m and density  $\rho=954.6$ kg/m$^3$. We set the pressure at the plate to the atmospheric pressure $p_1=101325$ Pa. The volume versus the curvature radius at the apex $R$ is shown in Fig.~\ref{fig:6}a and the pressure $p$, the energy $E$ and the angle $\alpha$ are shown versus $V$ in Figs.~\ref{fig:6}b, c and d, for $g=0$ and $g=9.8$ m/s$^2$. As expected, the effect of gravity in the different variables for small volumes is almost negligible.  For $g=0$, the contact angle increases monotonically with $%
V$ but the presence of a gravitational field induces a maximum angle $%
\alpha^{\max}=114.16^\circ$ after which it starts to decrease. This $\alpha^{\max}$ introduces
a condition on the wettability. For very small volumes, $\alpha\simeq 0$ and then $\theta\simeq180^\circ$  which
is necessarily larger than the advancing angle of the hydrophobic patch $%
\theta_2^{adv}$. Then, the contact line cannot be pinned at the patch. Therefore, the
first bubble that can be pinned should have an angle $\alpha=\pi-
\theta_2^{adv}$. Small bubbles are formed in the patch, moving and
coalescing until reaching a volume able to sustain an angle $%
\alpha=\pi-\theta_2^{adv}$ and form a single bubble. Then, the bubble is pinned
and the angle varies with the volume having a maximum  $\alpha^{\max}=\pi-\theta^{\min}$. If $\theta^{\min}<\theta_1^{rec}$ where $%
\theta_1^{rec}$ is the receding angle of the hydrophilic substrate, the liquid
dewets and the bubble base will grow beyond the patch.

Clearly large $\theta_2^{\rm rec}$ and small $\theta_1^{\rm rec}$ enhance boiling. In order to choose $r$, one should keep in mind that boiling occurs mainly at the contact line: a circular wedge of liquid phase in contact with the super heated substrate and with the vapour phase. A first guess is to choose  $r$ as small as possible so as to maximize the tip of the wedge length per substrate area.
Therefore a radius equal to the critical nucleation radius on the $\theta_2$ substrate at $T_{onb}$, namely $r_{onb}$, see Eq. \ref{ronb} .
As the bubble grows, the total tip of the wedge length decreases, so that heat transfer decreases. In order to maximize heat transfer, the bubble should detach as soon as possible. 
Bubble detachment requires gravity, therefore $r$ larger or of the order of the capillary length. A guide is to consider how detachment occurs for a uniform (un-patterned) substrate with $\theta_2^{\rm rec}$, modelled by Neumann b.c.. For this case, we can compute the maximum of the bubble contact radius before detachment, call it $r_{det}$. Now $r$ should be chosen in $(r_{onb},r_{det})$, actually smaller but of the order of $r_{det}$ so as to avoid depinning.
Then we can compute $r^{\rm opt}$ as minimizing the time to detachment, keeping in mind the pinning condition  $\theta>\theta_1^{\rm rec}$.
On each patch a bubble of contact angle $\pi-\theta_2$ (minimal bubble covering the patch) is left from the previous bubble and grows. If the patch radius is too small the bubble will eventually depin, and the benefit of patterning is lost. Otherwise it will detach without depinning. The instantaneous heat transfer is proportional to the molar rate of vaporization, which is proportional to $(T-373)$ K and to the contact line length, $2\pi r$ per patch, or proportional to $1/r$ per unit area of substrate because the number of patches is proportional to the inverse of the patch area $\pi r^2$. Therefore the optimal $r$, maximising heat transfer, is the smallest $r$ for which bubble detachment occurs before depinning. 
For each $(\theta_1,\theta_2)$ we can compute 'à la Longuet Higgins' $r^{\rm opt}$ as the patch radius $r$ such that, as the volume increases up to $V_{\max}$, the bubble contact angle increases up to the depinning angle $\pi-\theta_1$. Then, using this $r^{\rm opt}$, we can compute the mean rate of heat transfer for each $(\theta_1,\theta_2)$ .

\subsection{Effect of the patch radius} 

\begin{figure}[htp!] 
 \centering
 \includegraphics[width=\textwidth]{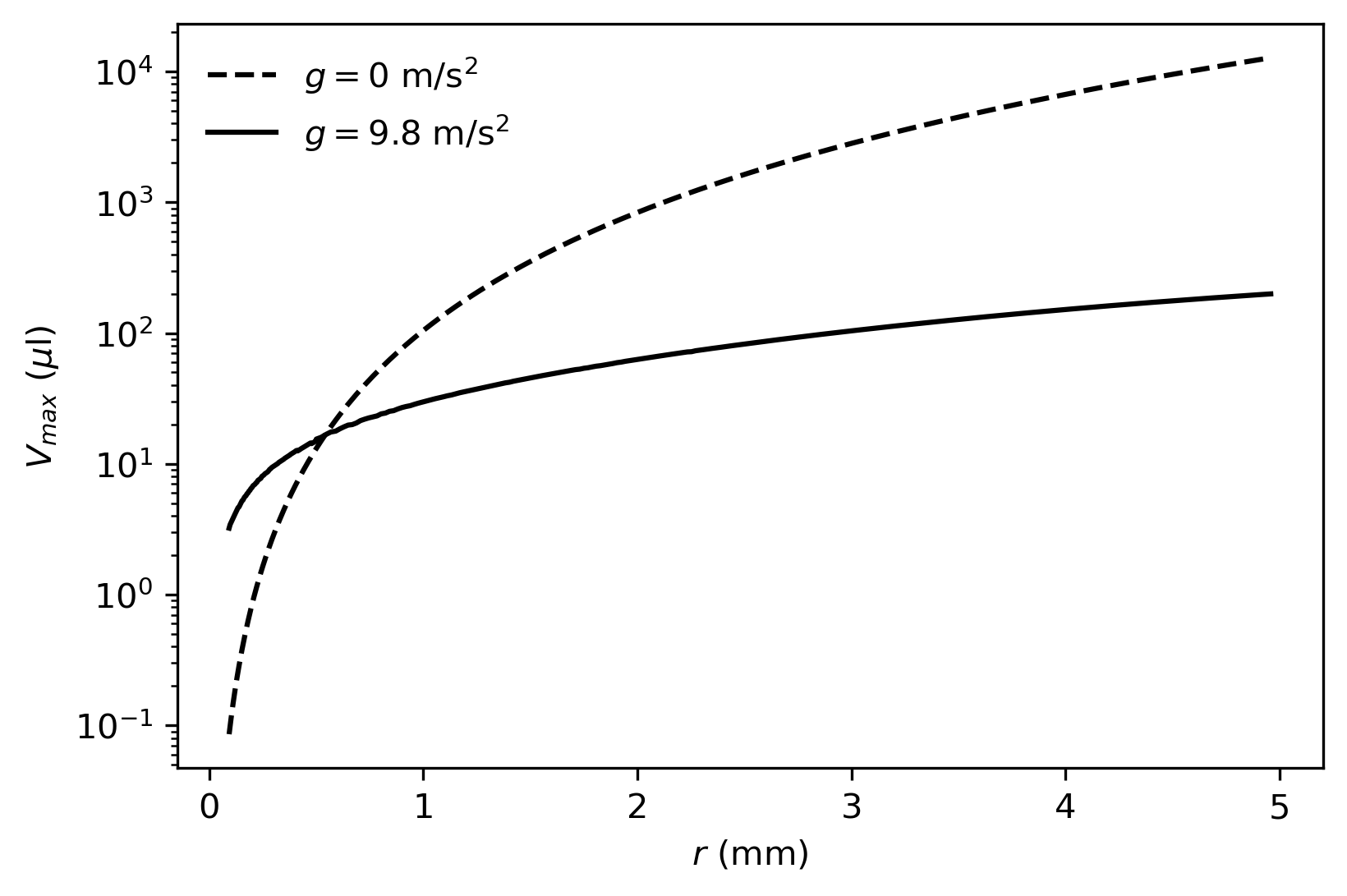} 
 \caption{Maximum volume versus $r$ before depinning ($g=0$ m/s$^2$),  or before instability ($g=9.8$ m/s$^2$), with $\theta_1^{rec}$=$20^\circ$, $\theta_2^{rec}=135^\circ$.}\label{fig:8}
\end{figure}

In Fig. \ref{fig:8} we show the maximum volume of a bubble $V_{\max}$ that can be
trapped in a patch of radius $r$ before instability or depinning (note that the volume axis is in
log scale). Depinning for $g=0$ depends on the
value of the receding contact angle. In the current example we consider $\theta_1^{rec}=20^\circ$. This value cannot be reached in the case $g=9.8$
m/s$^2$ where we have seen that the minimum contact angle from the water
side was $56.2^\circ$. The bubble will then grow until $V_{\max}$ defined
as the maximum volume for which a solution to the LH equations exists, \cite{LH91}.

If we want to determine the maximum volume of bubbles per unit substrate  area,
we need to divide this maximum volume by O($r_{\max}^2$) where $r_{\max}$ is half the
maximum width that the bubble achieves during its evolution until $%
V_{\max} $. Indeed, the density of patches should be chosen so as to avoid coalescence of neighboring bubbles, therefore a density of order $r_{\max}^{-2}$.

 \begin{figure}[htp!] 
 \centering
\includegraphics[width=\textwidth]{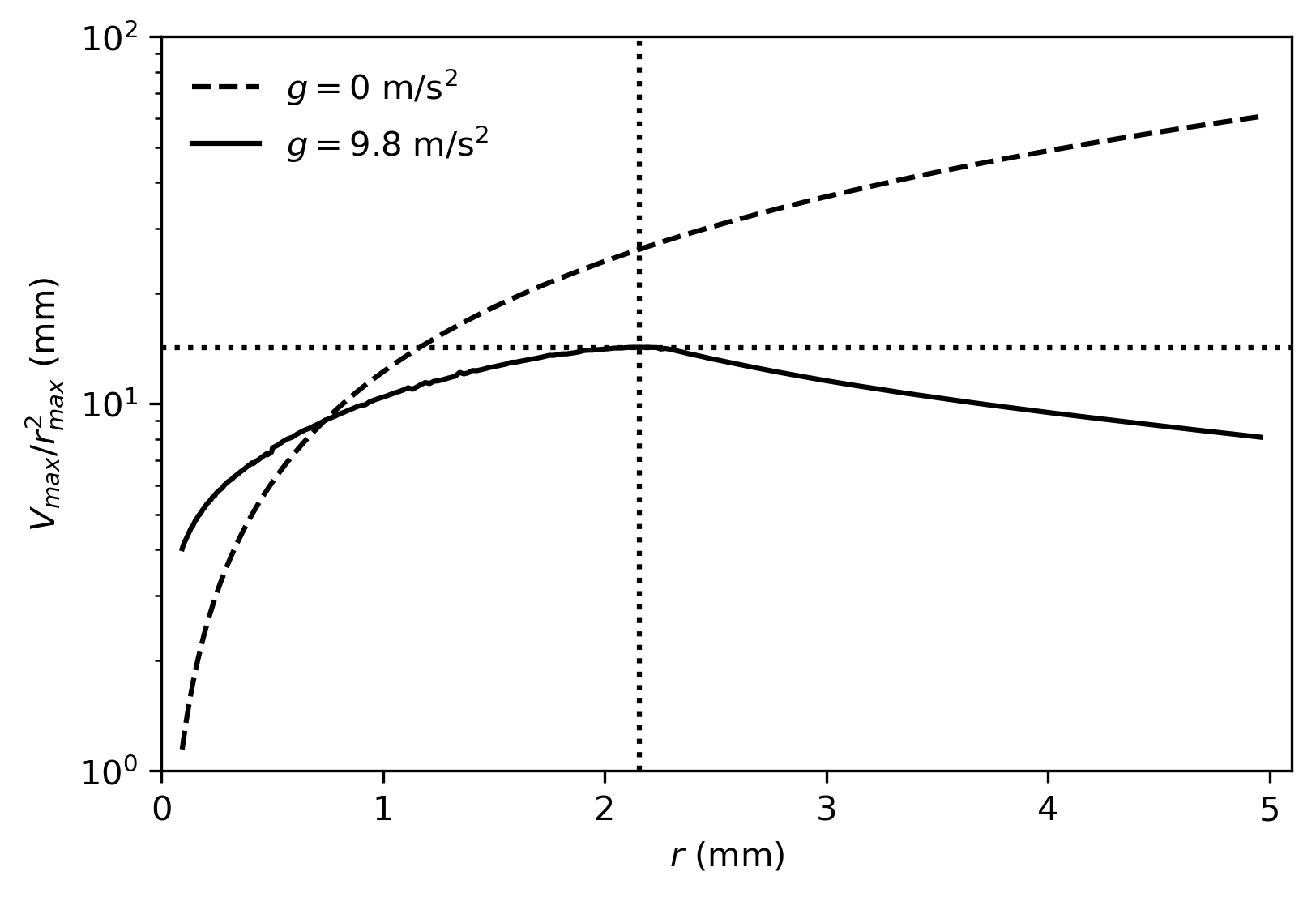}
 \caption{Maximum volume per unit area versus $r$ before depinning ($g=0$ m/s$^2$),  or before instability ($g=9.8$ m/s$^2$), with $\theta_1^{rec}$=$20^\circ$, $\theta_2^{rec}=135^\circ$.}\label{fig:9}
\end{figure} 

As shown in Fig.~\ref{fig:9}, for this particular example with $\theta_2^{rec}=135^\circ$, gravity induces a maximum in $V_{\max}/r_{\max}^2$ located at $%
r^{opt}=2.16$ mm. Therefore, we can maximize the total volume of
bubbles trapped by introducing patches with this radius.

\begin{figure}[htp!] 
 \centering
\includegraphics[width=\textwidth]{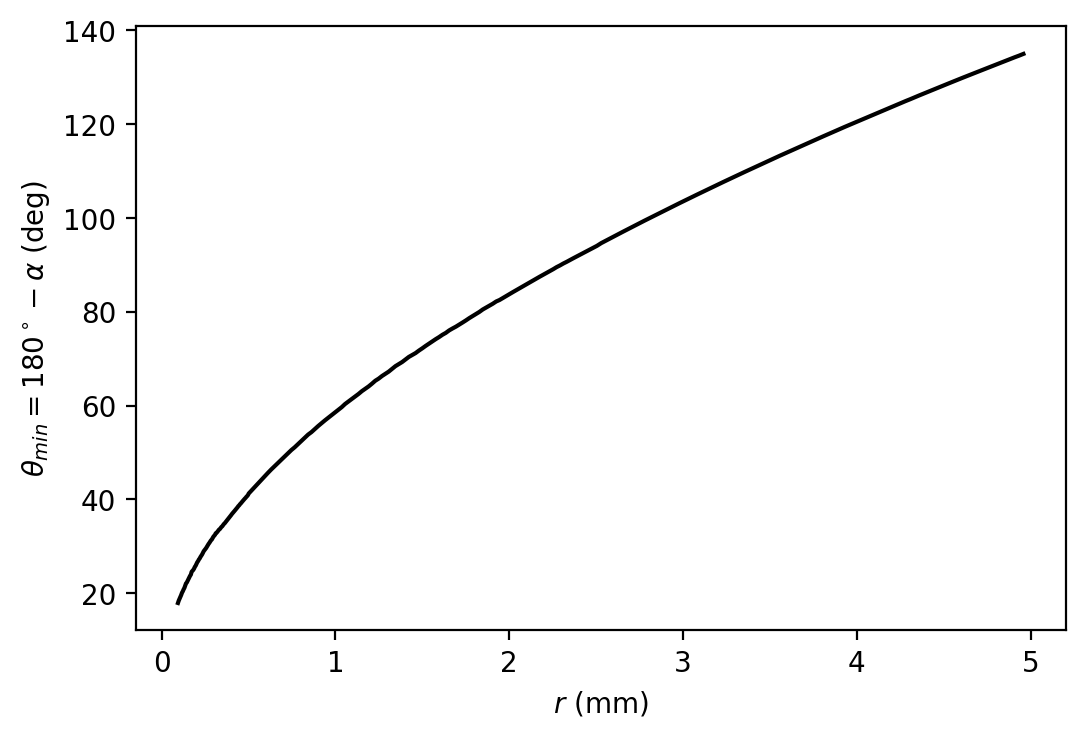}
 \caption{Minimum water contact angle versus patch radius $r$. }\label{fig:10}
\end{figure}

In Fig.~\ref{fig:10} we show the  minimum contact angle $\theta_{\min}=180^\circ-\alpha_{\max}$  achieved by
the bubble during its evolution as a function of $r$, for $g=9.8$. Here, we skip the comparison with $g=0$ where the maximum contact angle should be $180^\circ$.

For $r^{opt}$ the maximum angle is found to be $87.1^\circ$.
Therefore, if we consider the bubble evolution on an patch with optimal radius as a
series of stationary states, a hydrophobic patch with $\theta_2>%
\theta_{\min}^{opt}$ should be enough to pin the bubble until it reaches
its maximum value. Here, $\theta_{\min}^{opt}$ is the value of $\theta_{\min}$ at $r^{opt}$.

\bigskip

{\bf ACKNOWLEDGMENTS}

Special thanks to M. Marengo for stimulating
discussions. The authors also thank the European
Space Agency (ESA), France and the Belgian Federal
Science Policy (BELSPO) for their support in the
framework of the PRODEX Programme. This research
was partially funded by FNRS and Région Wallonne.

\bibliographystyle{plain}
\bibliography{main.bbl}

\end{document}